\documentclass[11pt]{article}
\usepackage{amsbsy}
\usepackage{amsmath}
\usepackage{amssymb}
\usepackage{graphicx}
\usepackage{dcolumn}
\usepackage{psfrag}
\usepackage{color}

\begin{document}

\title{\Large\bf Collisional Aggregation due to Turbulence}
\author{{\it  Alain Pumir $^1$ and
Michael Wilkinson$^2$}
\\[3 mm]
\normalsize($^1$)
Laboratoire de Physics, Ecole Normale Sup\'erieure de Lyon, CNRS\\
 and Universit\'e de Lyon, F-69007, Lyon, France$^1$\\
($^2$) Department of Mathematics and Statistics,
The Open University, \\Walton Hall, Milton Keynes, MK7 6AA, England
}
\vspace{3mm}
\date{}
\maketitle
\par
\vspace{2cm}

 {\large\bf Abstract}
 Collisions between particles suspended in a fluid play an important 
role in many physical processes. 
As an example, 
collisions of microscopic water droplets in clouds are a necessary
step in the production of macroscopic raindrops. Collisions of dust grains
are also conjectured to be important for planet formation
in the gas surrounding young stars, and also to play a role in the
dynamics of sand storms. 
In these processes, collisions are favoured by fast turbulent motions.
Here we review recent advances in the 
understanding of collisional aggregation due to turbulence. We discuss the role 
of fractal clustering of particles, and caustic singularities of their velocities. We
also discuss limitations of the Smoluchowski equation for modelling these 
processes. 
These advances lead to a semi-quantitative understanding on the 
influence of turbulence on collision rates, and point to deficiencies
in the current understanding of rainfall and planet
formation.

\par
\vspace{1.6cm}
\newpage
\newpage

\section{Introduction}
\label{sec: 1}

Fluids which are encountered in nature often carry a suspension of small particles~\cite{Bala+10}, which could
be dust grains, small liquid droplets or objects of more complicated shape,
such as ice crystals~\cite{Pru+97,Shaw03}, 
or even living organisms~\cite{Goldstein15}.
Their presence is typically evident from the optical behaviour of the fluid due to scattering
of light by the suspended particles, and this is the reason why the atmosphere or the sea are
often opaque or cloudy. In clouds, the absorption of short-wavelength radiation from
the sun, and of long-wavelength radiation from the earth, play an essential,
yet incompletely understood, role in the energy budget of the 
planet~\cite{Trenberth:2009}. 
By scavenging of aerosol particles and of other gases, 
water droplets contribute in an essential way to the dynamics of the atmosphere
and of the climate. Reliably modelling the distribution of particle sizes in 
suspensions is therefore essential to understand these processes.

As well as influencing optical properties, the stability of suspended particles in large
bodies of fluid can have important implications. The formation
of rain drops by the coalescence of vast numbers
of microscopic water droplets which make up atmospheric clouds, is 
one of the central questions facing cloud microphysics, which continues
to receive much attention~\cite{Shaw03,Grabo+13}.
Also, the standard model for formation of planets depends upon the 
collision and coalescence of dust grains in the atmosphere around a young star \cite{Safranov:69}. Explaining the existence of our planet, and
the weather phenomena that make it habitable, depends upon the instability of 
aerosol suspensions to collisions.

Attempts to make a quantitative theory for rain initiation or planet 
formation, however, run into difficulties. 
The collision rates of the atmospheric aerosol particles, resulting from
differential settling rates of water droplets of different sizes or from Brownian motion,
appear to be insufficient to explain the rapid onset of rain from many types of cloud.
Similarly, it is difficult to argue that the collision rate of dust grains is adequate to
explain planet formation.

One possible route to resolving these problems involves 
turbulence~\cite{ST56,Vol+80}. Turbulent motion is a
robust phenomenon, which occurs in many situations where a large body of 
fluid is in motion.
It is well known that the dispersion of small particles in a turbulent 
environment is
much faster than can be achieved by molecular 
diffusion~\cite{Taylor:21}, which suggests
that turbulence could also greatly enhance collision rates.
This argument  implies that turbulence
could also play an essential role 
in the coagulation process of particles.

The mechanisms whereby turbulence can enhance the collision rate have only become clear
in the last few years. This review will explain the current understanding, which is
forming a coherent and internally consistent picture, supported by numerical experiments
on accurate simulations of turbulent flow. The theoretical picture of the collision rate
proved to be multifaceted, involving concepts from dynamical systems theory, fractal geometry,
stochastic processes and optics, as well as results from fluid dynamics and the theory of
turbulence.

\begin{figure}
\includegraphics[width=0.55\textwidth]{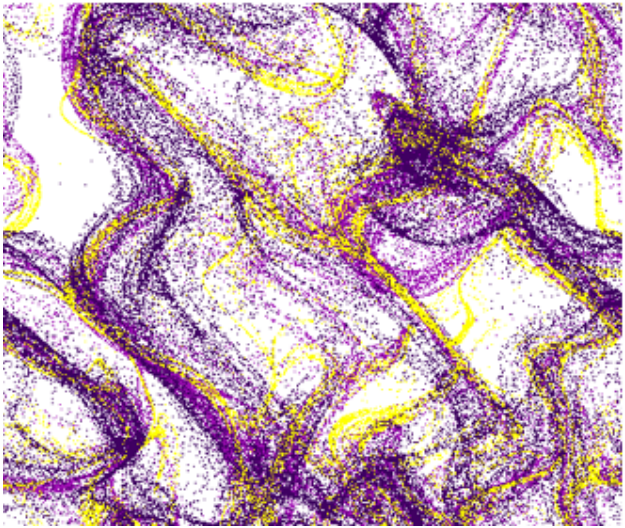}
\caption{
\label{fig: 1.1}
Particles in a turbulent flow can show pronounced clustering, 
which samples a fractal measure. The image is a two-dimensional
simulation of particles in an incompressible flow. The simulation 
includes particles with three different masses, shown in different colors.
}
\end{figure}

While some aspects of quantifying collision rates of small
particles in turbulent flows may require further 
work, it seems that the underlying physical principles 
are now qualitatively well understood.
The knowledge gained 
is expected to lead to a deeper understanding of collective behaviours in 
turbulent suspensions. It
remains to apply this knowledge to significant problems such as 
explaining rainfall, planet formation, and properties
of particle-laden turbulent flows such as sandstorms or powder-snow 
avalanches~\cite{Elghobashi:94}. At this level, 
fundamental problems remain outstanding. 
The general framework 
originally developed by Smoluchowski~\cite{Smol:17} to describe coagulation
processes in suspensions of Brownian particles, based on a 
mean-field approximation, may appear as an enticing
starting point. The rapid increase of the collision rate when the 
size of the particle increases, as it happens in the case of settling droplets 
in 
a cloud, can lead to runaway growth, which is a feature
of the formation of raindrops from microscopic water droplets. 
This phenomenon  is known as gelation in the polymer physics
literature. Surprisingly, it
has been shown that when gelation is modelled using the Smoluchowski
equation, the time required for the gelation transition may be strictly equal 
to zero~\cite{K+10}.
This instantaneous gelation is clearly unphysical;
it implies that 
mean-field descriptions based on the Smoluchowski approach have to be 
applied with caution. 

\begin{figure}
\includegraphics[width=0.9\textwidth]{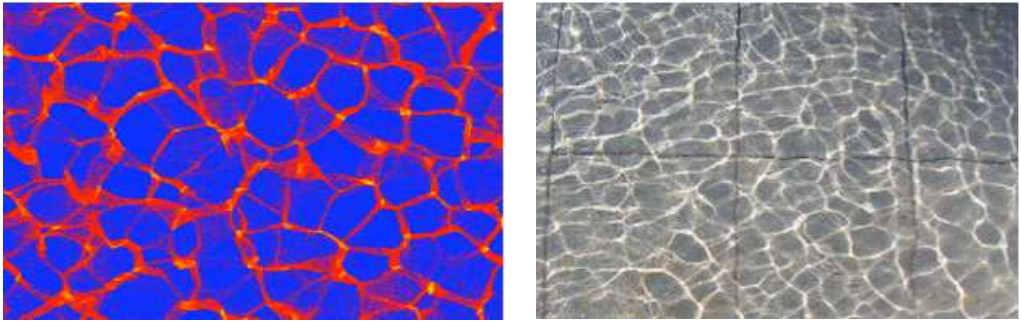}
\caption{
\label{fig: 1.2}
{\bf a} The distribution of particles in a turbulent flow can show 
singularities which result from the projection of folded 
manifolds in phase space: this is a simulation 
of a two-dimensional compressible flow, taken from \cite{WM05}. 
{\bf b} The singularities are termed 
caustics, because they have the same structure as singularities
in optics. The image (a photograph by Amanda W.  Peet) 
shows a pattern of caustics on the bottom of a swimming 
pool, resulting from partial focussing of sunlight due to curvature of the water surface 
by irregular waves. The two images are remarkably similar.
}
\end{figure}

We provide here
a critical discussion of results concerning the rate of rain drop and 
planet formation, which shows that while turbulence can
dramatically increase the collision rate, it is not entirely
clear whether it is sufficient to explain 
the coagulation rates observed in nature, and
alternative physical explanations may be required. 

Over the past few years, the subject of collisions induced by turbulence
has received a surge of interest, and as such, it has been the subject 
of several other review articles, focusing mostly on meteorological 
applications~\cite{Bod+10,Devenish+12,Grabo+13}. This review is 
more focused
on the lessons learned from a general fundamental physics
perspective. It is our belief
that some of the results shown here will be relevant in other fields of
physics, and thus of lasting interest, beyond the specific motivations of the
original study. 

This article is organised in the following way. In section
\ref{sec: 2}, we review elementary material concerning the motion of particles in a 
fluid flow, and the collision rate in a suspension, independently
of the precise nature of the flow. The more specific aspects of turbulent
flows, necessary for the purpose of this review, are discussed in 
section \ref{sec: 3}. Section \ref{sec: 4} will introduce various 
mechanisms whereby turbulence can influence the collision rate, 
including particle clustering and the effects of caustics forming in the 
phase space of the suspended particles. 
These diverse contributions to the collision rate are synthesised 
into a unified approximation scheme in section \ref{sec: 8}, which is 
shown to  provide a very accurate description of 
the results of numerical simulation.
Having discussed the determination for collision rates we turn to 
discussion of applications. Section \ref{sec: 9} considers the 
question of whether aerosols undergo a gelation transition, and whether
the Smoluchowski equation provides an appropriate description. Sections 
\ref{sec: 10} and \ref{sec: 11} consider applications to rainfall and planet 
formation. In both cases we argue that the insights from considering 
turbulent enhancement of collision rates do not appear to be sufficient
to resolve all of the problems with understanding these processes.
The prospects for further developments are considered in section \ref{sec: 12}, 
which is our conclusion. 
A detailed analysis of both fractal clustering and caustics,
which are crucial to the determination of the collision rate, is
provided in the case of a solvable one-dimensional model in Appendix A
(Section \ref{sec: 6}).
 
\section{Definitions and equations}
\label{sec: 2}

\subsection{Particle motion in a fluid flow}
\label{sec: 2.1}

The determination of the motion of particles, even in simple (laminar) flow 
configurations, is a difficult task. In the applications we have in mind, 
particles have small sizes, and this leads to a solvable problem. 
In this limit, the flow can be 
approximated as constant over a domain much larger than the particle, and 
the Reynolds number of the particle is so small that the nonlinear
term in the Navier-Stokes equations drops out, so the problem can
be explicitly solved~\cite{MaxRil83,Gat83}. 
The equations of motion involve several terms, which result from the
viscous drag of the particle, the gravitational settling, pressure effect,
the so-called added mass, and 
an history (Basset-Boussinesq) force.

Rain drops in a cloud or particles in the 
interstellar  medium 
have densities, $\rho_{\rm p}$, which are much larger than the fluid
density, $\rho_{\rm f}$: $\rho_{\rm p}/\rho_{\rm f} \gg 1$. The 
resulting inertia of the particles may be large enough to prevent them
from exactly following the flow. 
In this limit, 
it has been demonstrated~\cite{Elgho92}
that the two dominant
forces on small particles are due to viscous drag, which
causes the particle velocity to relax towards that of the fluid, and
to gravitational settling.
The equations of motion for small spherical particles of radius $a$ are
determined by Stokes' formula, so that the equation of motion is
\begin{equation}
\label{eq: 2.1}
\dot{\mbox{\boldmath$r$}}=\mbox{\boldmath$v$}
\ ,\ \ \
\dot{\mbox{\boldmath$v$}}=\frac{1}{\tau_{\rm p}}[\mbox{\boldmath$u$}(\mbox{\boldmath$r$},t)-\mbox{\boldmath$v$}] + \mbox{\boldmath$g$}
\end{equation}
where
\begin{equation}
\label{eq: 2.2}
\tau_{\rm p}=\frac{2}{9}\frac{a^2}{\nu}\frac{\rho_{\rm p}}{\rho_{\rm f}}
\end{equation}
is the particle relaxation time, determined from Stokes' formula for the
drag on a moving sphere (in (\ref{eq: 2.2}), $\nu$ is the kinematic 
viscosity). The equations of motion (\ref{eq: 2.1},\ref{eq: 2.2}) are
valid only 
in the limit where the suspended particles are very small and very dense:
$\rho_{\rm p}/\rho_{\rm f} \gg 1$. When the ratio 
$\rho_{\rm p}/\rho_{\rm f}$ 
is smaller than $\sim 10$, 
the history force becomes important~\cite{DT:11}.
In cases where a droplet moves through a fluid which has a comparable
viscosity, equation (\ref{eq: 2.2}) must be modified \cite{Ryb11,Had11}.

In astrophysical applications, the gas phase may have extremely 
low density, so that the size of the aerosol particles may be much smaller than the
mean free path of the gas. In this case equation (\ref{eq: 2.1}) still provides 
an accurate description of the motion, but the relaxation time $\tau_{\rm p}$ 
is given by a different expression:
\begin{equation}
\label{eq: 2.3}
\tau_{\rm p}=K_{\rm E} \frac{a \rho_{\rm p}}{c_{\rm s}\rho_{\rm g}}
\end{equation}
where $c_{\rm s}$ is the velocity of sound in the gas 
with density $\rho_{\rm g}$, and $K_{\rm E}$ is a 
dimensionless constant which is of order unity~\cite{Eps23}. 
The value of $K_{\rm E}$ depends on 
assumptions 
about the energy transferred when a gas molecule collides with the surface
of the particle. 

Describing the motion of larger particles in a turbulent flow 
remains an outstanding challenge, whose numerical and experimental study 
is at an early stage \cite{Zimm+11,Klein+13,Naso:10,Lucci+10,Homann+13}.

\subsection{Collision rates}
\label{sec: 2.2}

In principle, the calculation of the collision rate appears to be a 
straightforward exercise in elementary kinetic theory \cite{Boltz}, 
but in practice it is a surprisingly complex problem.
Before discussing the details, it is necessary to consider some
fundamental definitions.

Our objective is to quantify the rate $R$ for collision of 
a given particle with any other particle in the suspension. This quantity 
is expected to be a function of the radius $a$ of the particle concerned, 
so our objective is to calculate $R(a)$, which has dimensions 
of inverse time: $[R]={\rm T}^{-1}$. The rate of collision is expected
to be proportional to the number density of particles in the suspension. 
Let  $N(a) ~ \delta a$ be the number density of spherical particles
with radius in the small interval $[a,a+\delta a]$. The rate of collisions 
may be written
\begin{equation}
\label{eq: 2.2.0}
R(a)=\int_0^\infty {\rm d}a'\ \Gamma (a,a')\, N(a')
\end{equation}
where $\Gamma(a,a')$ is a characteristic property of the fluid 
motion, which is termed the collision kernel. Thus, the specific 
problem of determining the collision rate $R(a)$ in a given 
suspension is solved by determining the collision kernel 
$\Gamma(a,a')$, which is independent of the particle density, 
and then applying equation (\ref{eq: 2.2.0}).

The collision kernel may be expressed in terms of other  variables, 
for example 
when we formulate the Smoluchowski equations in section \ref{sec: 9}
it will be useful to represent the particle size distribution in term of masses,
so that the number density of particles in the interval $[m,m+\delta m]$ is 
$\bar N(m)\,\delta m$. In this case equation (\ref{eq: 2.2.0}) is 
replaced by
\begin{equation}
\label{eq: 2.2.0a}
R(m)=\int_0^\infty {\rm d}m'\ K(m,m')\, \bar N(m')
\end{equation}
where $K(m,m')$ is collision kernel expressed in terms of mass.
 
In a  monodisperse
suspension, where the particles all have approximately the same radius $a$, and where their 
number density is $n$, equation (\ref{eq: 2.2.0}) reduces to:
\begin{equation}
\label{eq: 2.2.0b}
R=\Gamma(a,a)\, n
\end{equation}
where 
\begin{equation}
\label{eq: 2.2.0c}
n =\int_0^\infty {\rm d}a\  N(a)=\int_0^\infty {\rm d}m\ \bar N(m)
\end{equation}
is the number density of particles.

\begin{figure}
\includegraphics[width=0.6\textwidth]{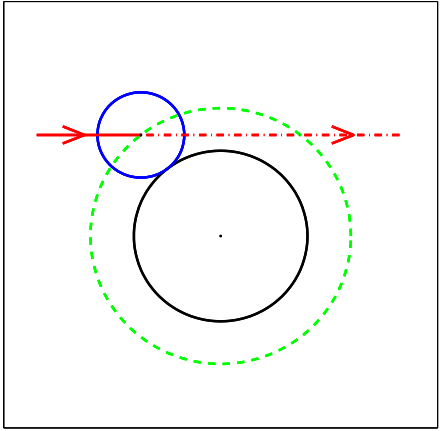}
\caption{
\label{fig: 2.1}
The rate of collision between particles of radius $a_1$ and 
$a_2$ is determined by integrating 
the relative velocity over a spherical surface of radius $a_1+a_2$, 
indicated by a dotted line.
}
\end{figure}

The collision rate between particles, which are assumed to be spherical, 
is the rate at which the separation vector between the centres of
the particles cross a sphere of radius $a_1 + a_2$, as 
illustrated in figure \ref{fig: 2.1}.
Thus, the collision kernel is expected to be proportional to the area
of the spherical surface, $4\pi (a_1+a_2)^2$. 
It is also proportional to a suitably defined average of the relative 
velocity of particles on the sphere of radius $a_1+a_2$, which will
be denoted by $\langle |\Delta v|\rangle$. Also, if the particles have a 
tendency to cluster together, the collision kernel would be proportional
to the two-point correlation function ${\cal C}(r)$ for particles with separation
$r=a_1+a_2$. Putting these factors together, the collision kernel can be
written in the form 
\begin{equation}
\label{eq: 2.2.1b}
\Gamma_{1,2} = \frac{1}{2}\, 4\pi\,(a_1+a_2)^2\,{\cal C}(a_1+a_2)\,\langle |\Delta v|\rangle
\ .
\end{equation}
The factor $\frac{1}{2}$ in (\ref{eq: 2.2.1b}) is included 
because only half of the particles crossing the surface are travelling inwards,
but this is really just a matter of convention in the definition of $\langle |\Delta v|\rangle$.
With a suitable definition of $\langle |\Delta v|\rangle$, this can be
presented as an exact equation, see \cite{Sundaram:97,Wang:00}.

\begin{figure}
\includegraphics[width=0.6\textwidth]{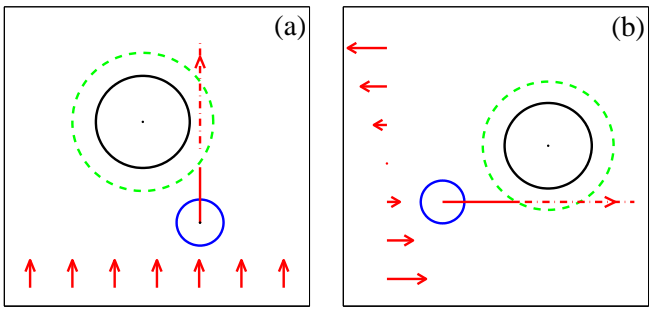}
\caption{
\label{fig: 2.2}
The relative velocity of particles might be independent of their
separation, illustrated in (a) for a larger particle overtaking a smaller 
one as they both fall under gravity. Alternatively, the relative velocity 
may be proportional to their velocity gradient: (b) shows a collision 
induced by a simple shear
flow (in both cases the arrows indicate the velocity of the smaller 
particle relative to the larger one).
}
\end{figure}

The collision kernel in (\ref{eq: 2.2.1b}) depends on the  
mean relative velocity of the particles, $\langle | \Delta v | \rangle$,
which is difficult to calculate.
The complex problem of determining the relative velocity between small
particles is simply tractable in two limits, relevant to the problem studied 
here. One case is where the relative velocities might be approximately 
independent of the separation of the particles, as illustrated in 
figure \ref{fig: 2.2}(a).
This  is only possible when the particles are able to move relative
to the fluid. The other limiting case is where the particles are 
advected with the fluid. For very small advected particles, their relative 
velocity at collision $\Delta \mbox{\boldmath$v$}$ 
is proportional to their separation $\Delta \mbox{\boldmath$x$}$:
\begin{equation}
\label{eq: 2.2.1a}
\Delta \mbox{\boldmath$v$}={\bf A}\,\Delta \mbox{\boldmath$x$}
\end{equation}
where ${\bf A}$ is the matrix of velocity gradients, with elements 
$A_{ij}=\frac{\partial u_i}{\partial x_j}$. 

The approach discussed here is purely geometric, and misses important effects. When
two particles get close to each other, the fluid trapped between them
leads to a lubrication film, which potentially significantly reduce the 
collision rates. This reduction 
can be taken into account by introducing a 
collision efficiency~\cite{Pru+97,Mas57}. It will be discussed in 
Section~\ref{sec: 9.1}.

In the following, we discuss three useful examples of physical situations,
amenable to an explicit determination of the collision kernel.

\subsubsection{Collision in a gas of particles with a Gaussian distribution of velocity}
\label{sec: 2.2.1}

For future reference, we estimate here the collision rate between particles
spatially uniformly distributed in a fluid, with {\em statistically 
independent} Gaussian (Maxwellian)
distribution of velocities, such that $\langle \mbox{\boldmath$v$} \rangle = 0$
and a variance $\langle \mbox{\boldmath$v$}^2 \rangle_{a}$, which
depends {\it a-priori} on the radius $a$ of the particles.
The assumption
that particles are uniformly distributed ensures that ${\cal C}(a_1 + a_2) = 1$. 
Assuming that the distribution of the velocity for particle $i$ is
$P(\mbox{\boldmath$v$}_i) = 
(\frac{3}{2\pi \langle \mbox{\boldmath$v$}^2 \rangle_{a_i}} )^{3/2}
\exp[- 3{\mbox{\boldmath$v$}_i}^2/ ( 2 \langle \mbox{\boldmath$v$}^2 \rangle_{a_i}) ]$,
an elementary calculation shows that the average velocity required in
(\ref{eq: 2.2.1b}) is 
$\langle | \Delta v_r | \rangle = [\frac{2}{\pi} ( \langle \mbox{\boldmath$v$}^2 \rangle_{a_1} 
+ \langle \mbox{\boldmath$v$}^2 \rangle_{a_2} ) ]^{1/2} $.
This leads to the following collision rate in a suspensions of 
particles with a Gaussian distribution of velocities:
\begin{equation}
\Gamma_{1,2} = (\frac{8 \pi}{3})^{1/2}(a_1 + a_2)^2 [ \langle \mbox{\boldmath$v$}^2 \rangle_{a_1}
 + \langle \mbox{\boldmath$v$}^2 \rangle_{a_2} ]^{1/2}
\end{equation}
consistent with classical estimates~\cite{LL_SP}.

\subsubsection{Collision between settling droplets in still air}
\label{sec: 2.2.2}

We consider now the case of polydisperse suspension of water 
droplets settling in still air. Integrating (\ref{eq: 2.1}), and 
taking into account the dependence on the size $a$ of $\tau_p$ leads to the 
following equations for the settling velocity $\mbox{\boldmath$v$}_s$:
\begin{equation}
\label{eq: 2.4}
\mbox{\boldmath$v$}_s = -v_s \hat{e}_z, ~ v_s = \kappa a^2
\ ,\ \ \
\kappa=\frac{2}{9}\frac{\rho_{\rm p}}{\rho_{\rm g}}\frac{g}{\nu}
\ .
\end{equation}
This assumes that the Reynolds number based upon the particle size 
is small, which is valid for 
the microscopic water droplets in clouds \cite{Mas57}. Equation (\ref{eq: 2.4})
shows a strong dependence of $v_s$ on $a$: large particles
settle much faster than smaller ones.
Consider the collision rate between particles of sizes $a_1$ and
$a_2$. In this case, the value of ${\cal C}(a_1 + a_2)$ reduces to $1$, 
and a simple 
calculation leads to 
$\langle | \Delta v_r | \rangle = \kappa |a_1^2 - a_2^2|/2$, 
which immediately leads to:
\begin{equation}
\label{eq: 2.5}
\Gamma_{1,2}  = 
\pi \kappa (a_1+a_2)^2 |a_1^2-a_2^2|
\ .
\end{equation}
This expression, which grows as a power $\propto a_1^4$, when $a_1\gg a_2$, 
i.e., with a power 
$4/3$ of the volume of the larger particles, implies that the collision rate of large
particles grows very rapidly as their size increases. This fast growth will have an
important consequence when studying the problem of coagulation, see Section
\ref{sec: 9}.

\subsubsection{Collision in a simple shear flow}
\label{subsubsec_adv_coll}

Whereas in the two previous examples, the motion of the fluid was not
playing any role, consider now a suspension of particles simply transported
in a simple (laminar) shear flow: $\mbox{\boldmath$u$} = S (y,0,0)$, so particles
are moving in the $x$-direction, with a velocity $\mbox{\boldmath$v$} = \mbox{\boldmath$u$}$, 
which depends only on its $y$-component (see figure \ref{fig: 2.2}). 
This problem was first treated by Smoluchowski \cite{Smol:17}.
Once again, assuming a uniform 
distribution of particles in the flow implies that ${\cal C}(a_1 + a_2)= 1$.
An elementary estimate of the $\Delta v_r$ leads to the following expression 
for $\langle | \Delta v_r | \rangle =  \frac{8}{3} |S|(a_1 + a_2 )  $,
which leads to the collision rate:
\begin{equation}
\Gamma_{1,2} = \frac{16 \pi}{3}\, |S|\, (a_1 + a_2)^3
\ .
\label{eq:coll_shear} 
\end{equation}
The collision rate, which  
is proportional to the shear rate, $S$, and the distance $(a_1 + a_2)$
to the {\it third} power, is characteristic of the collision rates
obtained as a result of shearing motion, at least when the particles follow 
the flow. In this case, the simple flow structure appears through the shear 
rate, $S$. 

\begin{figure}
\includegraphics[width=0.6\textwidth]{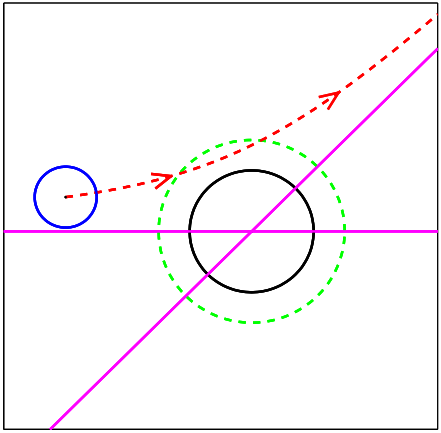}
\caption{
\label{fig: 2.3}
The velocity gradient is not usually a simple shear. 
A locally hyperbolic relative velocity also induces collisions.
}
\end{figure}

In turbulent flows, velocity differences are created at many scales, generating,
among other things, strong shearing motions on small scales. 
The velocity 
gradient in turbulence or other complex flow field is described by a 
real-valued $3\times 3$ matrix ${\bf A}$, which is generically not a 
simple shear. Other possibilities include the case of locally
hyperbolic flow, as illustrated in figure \ref{fig: 2.3}. 
To gain further insight on the statistics of 
$\langle | \Delta v| \rangle$ requires a more precise description of
the properties of the turbulent velocity field, which will be considered
in the next section.
 
\section{Elementary properties of turbulent flows} 
\label{sec: 3}

The enhancement of the collision rate by a simple flow, such as a shear 
flow considered in the previous subsection, provides the first hint 
concerning enhancement of collision rates by
turbulent flows. In fact, significant velocity gradients are ubiquitous in 
turbulent flows.
This section is aimed at providing an elementary discussion of the properties
of turbulence relevant to the present discussion.

In this review, we are concerned with turbulence in an incompressible,
Newtonian fluid, described by the Navier-Stokes equations:
\begin{eqnarray}
& \partial_t &\mbox{\boldmath$u$}  +  (\mbox{\boldmath$u$} \cdot \nabla) \mbox{\boldmath$u$} 
= - \frac{1}{\rho_{\rm f}} \nabla p
+ \nu \nabla^2 \mbox{\boldmath$u$} + \mbox{\boldmath$f$} \label{eq:3.1} \\
& \nabla \cdot & \mbox{\boldmath$u$} = 0
\label{eq:3.2}
\ .
\end{eqnarray}
where $p$ is pressure.
The viscous term dissipates kinetic energy in the fluid, so in the absence
of any forcing, $\mbox{\boldmath$f$} = 0$, the motion simply decays. The Reynolds number,
defined as ${\rm Re} = U L/\nu$, where $U$ and $L$ are the velocity 
and length scales at which the fluid is forced, measures the ratio between the 
nonlinear term,
$(\mbox{\boldmath$u$} \cdot \nabla) \mbox{\boldmath$u$}$ and the viscous (dissipative) term,
$\nu \nabla^2 \mbox{\boldmath$u$}$.
In turbulent flows, the Reynolds number is effectively a measure of the
intensity of turbulence: turbulence is expected to occur whenever the Reynolds number is very 
large: ${\rm Re} \gg 1$. Thus, the statement ${\rm Re} \gg 1$ effectively means that,
{\it at the forcing scale}, viscous dissipation does not play much role. 
Primarily because the kinematic viscosity $\nu$ is a small quantity, very large Reynolds 
numbers are common: for example, a convective instability of the atmosphere
can create a flow with ${\rm Re}>10^8$.

\subsection{Scales in a turbulent flow}
\label{sec: 3.1}

Turbulence is a notoriously difficult problem~\cite{Falk:06}
but we argue that most
of what we need to know can be surmised from dimensional considerations. 
Kolmogorov~\cite{Kolm:41a,Kolm:41b} introduced the powerful notion that
the small structures of the flow have very little
\lq memory' of how the flow was generated. He 
argued that only one quantity, 
the rate of kinetic energy dissipation per unit mass, 
$\epsilon$, is required to characterise 
fully developed steady and homogeneous turbulent motion over a wide
range of length scales (termed the \lq inertial range' in the turbulence 
literature). The quantity $\epsilon$, the power dissipated per unit
mass, and designated in the following in
short as \lq energy dissipation', is determined
by the manner in which the turbulence is generated, by means of large scale
flows with characteristic length scale $L$ and velocity scale $U$. Dimensional 
considerations imply that $\epsilon \sim U^3/L$. 
In the inertial range of length scales, where $\epsilon$ is the only relevant
parameter, statistics of the flow can be determined by dimensional analysis.  
For example, in the inertial range of separations, the variance of the velocity 
difference between two points depends only upon the separation 
$R=|\mbox{\boldmath$R$}|$ and the kinematic viscosity $\nu$. Dimensional
consistency then implies that
\begin{equation}
\label{eq: 3.3}
\langle [\mbox{\boldmath$u$}(\mbox{\boldmath$R$},t)-\mbox{\boldmath$u$}({\bf 0},t)]^2\rangle
=C\left(\epsilon |\mbox{\boldmath$R$}|\right)^{2/3}
\end{equation}
where the brackets in (\ref{eq: 3.3}) refer to an average over many
flow realizations, and $C$ is a universal dimensionless coefficient.

The turbulence generates successively finer scale eddies 
until the structures 
become so small that gradients increase, making the power dissipated 
per unit mass,
$\epsilon \sim \nu \mbox{\boldmath$u$} \cdot \nabla^2  \mbox{\boldmath$u$}$,  
significant.  
The smallest scale reached by the flow is 
the \lq Kolmogorov lengthscale', $\eta$,  with a characteristic time
scale known as the Kolmogorov timescale, $\tau_K$. 
These quantities  depend only on $\epsilon$ and $\nu$.  
Dimensional considerations then imply that
\begin{equation} 
\label{eq: 3.4}
\eta=(\nu^3/\epsilon)^{1/4}
\ ,\ \ \ 
\tau_K=(\nu/\epsilon)^{1/2}
\ .
\end{equation}

As finer scales are generated by the flow, it is expected that the 
statistical properties of the flow become homogeneous and 
isotropic~\cite{Kolm:41a,Kolm:41b}. 
In this case the kinetic energy
dissipation can be obtained directly from (\ref{eq:3.1}): 
\begin{equation}
\label{eq: 3.5}
\epsilon= \nu \sum_{i,j} \langle A_{ij}^2 \rangle
\end{equation}
where $A_{ij}$, the velocity gradient, is defined by (\ref{eq: 2.2.1a}).
This implies that the typical size of the velocity gradient is the inverse of the 
Kolmogorov time: $\partial u/\partial x\sim \tau_{\rm K}^{-1}$.

\subsection{Velocity gradient statistics }
\label{sec: 3.1a}

In turbulent flows, the relative motion of two small particles 
approaching (colliding with) each other is ultimately dominated by the
velocity gradient tensor, $\mathbf{A}$, as explained 
in Subsection~\ref{sec: 2.2}. For this reason, we briefly 
discuss some elementary statistical properties of the velocity gradient. 

The isotropy of the flow imposes that the tensor 
$\langle A_{ij}(\mbox{\boldmath$x$} ) A_{kl} (\mbox{\boldmath$x$} ) \rangle$ is expressible 
in terms of Kronecker $\delta$ tensors. Using 
the incompressibility condition ${\rm tr}( \mathbf{A}) = 0$, as well as 
relation 
(\ref{eq: 3.5}) leads to the following expression for 
$\langle A_{ij}(\mbox{\boldmath$x$} ) A_{kl} (\mbox{\boldmath$x$} ) \rangle$:
\begin{equation}
\label{eq: 3.6}
\langle A_{ij}(\mbox{\boldmath$x$}) A_{kl}( \mbox{\boldmath$x$} ) \rangle
= \frac{\epsilon}{ 30 \nu} \Bigl( 4 \delta_{ik} \delta_{jl} - \delta_{il}\delta_{jk} - \delta_{ij} \delta_{kl} \Bigr)
\ .
\end{equation}
Such an estimate is
crucial in establishing elementary results, such as the Saffman-Turner
collision rate, Eq.~(\ref{eq: 4.1.6}).  

As a particle is transported by the flow, the strain and the vorticity 
along its trajectory decorrelate with 
a correlation time of the order of $\tau_K$~\cite{Brunk98,PW11}. 

\subsection{The Stokes number }
\label{sec: 3.2}

Comparing the Kolmogorov time $\tau_{\rm K}$ with the response time
of the particles, $\tau_{\rm p}$, provides a way to quantify the 
effect of inertia. This motivates the definition of the Stokes number:
\begin{equation}
\label{eq: 2.8}
{\rm St}=\frac{\tau_{\rm p}}{\tau_{\rm K}}
\ .
\end{equation}

For ${\rm St}\ll 1$, particles are advected by the fluid, and collisions are
the result of the shear (the relative motion). 
When ${\rm St}\gg 1$, the inertia of the particles allows
them to move relative to the surrounding fluid, thus leading to entirely
different phenomena. The Stokes number is the single dimensionless
parameter which distinguishes different physical regimes of the collision 
process.

\subsection{Experimental and numerical investigations}
\label{sec: 3.3}

\subsubsection{Experimental studies}
\label{sec: 3.3.2}

Despite the vast literature devoted to the experimental 
investigation of turbulence, very little is known experimentally 
concerning collisions of particles suspended in turbulent 
flows.
 
The investigation of turbulent flows has rested for a long time on methods,
such as hot-wire anemometry, which provide only information on the velocity
and its spatial correlation function.
Over the past decade, new methods have been developed, based
on following particles in a turbulent flows using 
fast-imaging~\cite{LaPorta:01,Toschi-Bod:09}. This has
led to a wealth of new information on the motion of particles in a turbulent
flows~\cite{Toschi-Bod:09}. 
Although in principle feasible, detecting collisions between
small particles in a well-controlled laboratory flow has
so far not been possible. It is to be expected that this problem 
will be solved in a near future. Labelling liquid droplets with chemicals 
that produce an optically detectable reaction product when the droplets 
coalesce is a promising approach.

\subsubsection{Numerical investigations}
\label{sec: 3.3.1}

The difficulty in obtaining experimental results on collisions 
in turbulence makes numerical investigations an essential tool. 
The investigation of fundamental issues in turbulence rests on direct
integration of the Navier-Stokes equation, (\ref{eq:3.1}) and (\ref{eq:3.2}),
in a triply periodic domain (effectively a torus)~\cite{OrszagPatt,Kaneda:09,PK:11}. 
In such a configuration,
the Navier-Stokes can be efficiently integrated using pseudo-spectral methods,
based on a (truncated) Fourier series decompositions of the velocity
field $\mbox{\boldmath$u$}$. For efficiency purposes, the calculation of the nonlinear 
term $( \mbox{\boldmath$u$} \cdot \nabla) \mbox{\boldmath$u$}$ is 
carried out in real space, using
fast-Fourier methods to transform between real-space and Fourier representations.

Once a solution of the Navier-Stokes equations is determined, 
the motion
of particles, (\ref{eq: 2.1}), can be efficiently 
solved~\cite{YeungPope:88}, permitting
to detect collisions~\cite{Sundaram:96}. The available computer
resources permit to address most aspects of particle collisions.

\section{Effect of turbulence on the collision rate}
\label{sec: 4}

One may think that the effect of turbulence reduces to an effective
\lq rate of strain'. The result is more complicated, and much more interesting,
as discussed in this section. In the following we describe various effects 
which become significant as the Stokes number is increased.
In the following sections, we focus on the case of monodisperse 
suspensions, and discuss the collision rate $R$ per particle.

As well as influencing the settling of particles from a fluid suspension
by facilitating collisions and aggregation, turbulent motion can have a direct
effect upon the settling rate of heavy particles~\cite{Good+14}. 
These single-particle effects are usually less significant than the collisional 
processes which are the focus of this review.

\subsection{The Saffman-Turner limit}
\label{sec: 4.1}

Particles with a sufficiently small inertia (small Stokes numbers)
essentially follow the flow (independent of their shape or material density).
In this case, the velocity gradients generated by turbulence strongly
enhance the
relative motion between particles, 
therefore enhancing the chance of collisions, by the mechanism 
shown in the simple example treated in \ref{subsubsec_adv_coll}.
A formula derived by Saffman and Turner~\cite{ST56}, equation (\ref{eq: 4.1.6})
below, determines the collision rate in the small Stokes number case, in the case 
where the particles are spherical. 
Two small particles of radius $a$ advected by the flow 
collide if their separation falls below $2a$. In the case of a suspension of
particles uniformly distributed throughout the fluid with 
density $n$, the rate of collision $R$
for a single particle is obtained, in the spirit of (\ref{eq: 2.2.1b}),
by integrating 
the relative radial velocity $\Delta v_r$ over the surface of a sphere of 
radius $2a$:
\begin{equation}
\label{eq: 4.1.1}
R=\frac{1}{2}  n \int {\rm d}\mbox{\boldmath$S$}\cdot |\Delta v_r|
\ .
\end{equation}
In this case the relative velocity is determined by the linearisation of the flow
field, so that the relative motion of two particles may be described by 
a hyperbolic velocity
field, such as that illustrated in figure \ref{fig: 2.3}.
The factor of $\frac{1}{2}$ in (\ref{eq: 4.1.1}) is required because 
only half of the sphere where the 
relative velocity is negative contributes to the collision rate (the overall flux is zero, due to incompressibility).
If the particles 
are small compared to the Kolmogorov length of the flow, the relative velocity 
resulting from the action of the local velocity gradient $\mathbf{A}$
is given by Eq.~(\ref{eq: 2.2.1a}).
Using the explicit expression $A_{ij} = \partial u_i/\partial x_j$ leads
to the following expression for the velocity gradient: 
\begin{equation}
\label{eq: 4.1.3}
\langle |\Delta v_r| \rangle = 2a ~ \langle | \mathbf{n} \cdot \mathbf{A} \cdot \mathbf{n} | \rangle 
\end{equation}
where $\mathbf{n}$ is an arbitrary unit vector on the unit sphere. Because of the 
isotropy of the velocity field, we have
\begin{equation}
\label{eq: 4.1.3a}
\langle | \mathbf{n} \cdot \mathbf{A} \cdot \mathbf{n} |  \rangle 
= \langle \left\vert \frac{\partial u_x}{\partial x}\right\vert \rangle
\ .\end{equation}
This leads to:
\begin{equation}
\label{eq: 4.1.4}
R=2\pi (2a)^3  n  \langle \left\vert\frac{\partial u_x}{\partial x}\right\vert \rangle 
\ .
\end{equation}
In order to estimate the expectation value of the partial derivative 
$|\partial u_x/\partial x|$ we assume that the elements of ${\bf A}$ are 
Gaussian distributed, and note that the variance of $\partial u_x/\partial x$
is given by $\epsilon/15\nu$ (see equation (\ref{eq: 3.6})).
The averaged value  
is equal to the inverse of the Kolmogorov time scale, up to numerical 
prefactors~\cite{ST56}, leading to:
\begin{equation}
\label{eq: 4.1.6}
R= \sqrt{\frac{8\pi}{15}}\frac{n (2a)^3}{\tau_{\rm K}}
\ .
\end{equation}
This estimate of the collision rate is based 
on the only assumption that particles are uniformly distributed, and can
be easily extended to suspensions of particles with a dispersion of size~\cite{ST56}.

\subsection{Clustering in turbulent flows}
\label{sec: 4.2}

When the effects of inertia become sufficiently large, i.e., when ${\rm St}$
is large enough, other effects arise, and the collision rate
cannot be simply understood in terms of relative motion in the fluid.
At finite ${\rm St}$ it is known that particles can show a pronounced 
clustering.  We assume here that the 
local particle density remains small enough, to prevent feedback from the
particles on the flow.

\subsubsection{Clustering and the centrifuge effect}

Particle clustering is expected to increase 
the collision rate. 
The effect of clustering is often ascribed to particles being expelled 
from vortices by centrifugal action, as proposed by Maxey~\cite{Max87},
based on the following argument.
Integration of the equation of motion (\ref{eq: 2.1}) gives
\begin{equation}
\label{eq: 5.1.1}
\mbox{\boldmath$v$}(t)=\int_{-\infty}^t{\rm d}t'\ 
\mbox{\boldmath$u$}(\mbox{\boldmath$x$}(t'),t')\, \exp[-(t-t')/\tau_{\rm p}]
\ .
\end{equation}
The expression (\ref{eq: 5.1.1}) reduces, in the case where 
$\tau_{\rm p}$ is small, to:
\begin{equation}
\label{eq: 5.1.2}
\mbox{\boldmath$v$}=\mbox{\boldmath$u$}(\mbox{\boldmath$x$}(t),t)
-\tau_{\rm p}\frac{{\rm D}\mbox{\boldmath$u$}}{{\rm D}t}
(\mbox{\boldmath$x$}(t),t)
\end{equation}
where ${\rm D}/{\rm D}t=\partial/\partial t+\mbox{\boldmath$u$}\cdot 
\mbox{\boldmath$\nabla$}$. 
The important remark is that the velocity field $\mbox{\boldmath$v$}$, which 
transports the particles, differs in
an essential way from the flow velocity field, $\mbox{\boldmath$u$}$: $\mbox{\boldmath$v$}$
is effectively {\it compressible}, with a divergence
\begin{equation}
\label{eq: 5.1.3}
\mbox{\boldmath$\nabla$}\cdot \mbox{\boldmath$v$}
= -\tau_{\rm p} \sum_{ij}\left(\frac{\partial u_i}{\partial x_j} 
\frac{\partial u_j }{\partial x_i} \right)
=  -\tau_{\rm p}  {\rm tr}({\bf A}^2) 
\end{equation}
where ${\bf A}$ is the velocity gradient tensor. 
Expressing ${\bf A}={\bf E}+\mbox{\boldmath$\Omega$}$,
where ${\bf E}^{\rm T}={\bf E}$ is the strain rate tensor (symmetric) 
and $\mbox{\boldmath$\Omega$}^{\rm T}=-\mbox{\boldmath$\Omega$}$
is the vorticity tensor (antisymmetric) leads to: 
\begin{equation}
\label{eq: 5.1.4}
\mbox{\boldmath$\nabla$}\cdot \mbox{\boldmath$v$}
= -\tau_{\rm p}\left[ {\rm tr}({\bf E}^2)
-\frac{1}{2}\mbox{\boldmath$\omega$}\cdot \mbox{\boldmath$\omega$}\right]
\end{equation}
 where $\mbox{\boldmath$\omega$}$ is the axial vorticity vector corresponding
to the antisymmetric tensor $\mbox{\boldmath$\Omega$}$,
so that the particle flow is contracting in regions of high strain 
and expanding in regions of high vorticity. 
Numerical evidence of a negative correlation between particle 
density and vorticity when ${\rm St}\approx 1$, as predicted by the \lq centrifuge'  
argument has been found numerically \cite{Wan93}.
We stress that depends upon making the approximation that $\tau_{\rm p}$ is small.
Using (\ref{eq: 5.1.4}) provides to a method, valid in the limit of small 
Stokes number, to analyse clustering~\cite{Bal01,FFS02,FP04}.

\subsubsection{Clustering as a manifestation of a fractal distribution}

Figure \ref{fig: 1.1} illustrates the type of clustering phenomena 
which can occur in turbulent flows. The clustering effect is much stronger than
the simple \lq centrifuge effect' argument suggests, and the positions
where the greatest enhancement of density will occur are not 
predicted by that argument. In reality, the clusters have fractal properties, 
which should be quantified to determine their influence on collision rates.
One way of characterising the fractal distribution 
of particles is via the correlation dimension \cite{Gra+84,Ott02}. 
To define this, pick a particle at random,
and determine the number ${\cal N}(\epsilon)$ of other particles within a ball 
of radius $\epsilon$ 
centred on this particle. The set can be characterised by averaging 
${\cal N}(\epsilon)$. 
The quantity is related to the pair correlation function ${\cal C}(\epsilon)$ of 
the particle 
distribution:
\begin{equation}
\label{eq: 4.2.2}
{\cal C}(\epsilon)=\frac{1}{4\pi  n \epsilon^2}
\frac{{\rm d}\langle {\cal N}\rangle}{{\rm d}\epsilon}\bigg\vert_{\epsilon}
\ .
\end{equation}
If the particle positions sample a fractal measure, it is 
expected that $\langle {\cal N}\rangle(\epsilon)$ has a power-law
dependence:
\begin{equation}
\label{eq: 4.2.3}
\langle {\cal N}(\epsilon)\rangle\sim  n\eta^3\left(\frac{\epsilon}{\eta}\right) ^{D_2}
\end{equation}
where $D_2$ is the correlation dimension of the set. This implies that 
the pair correlation function is 
\begin{equation}
\label{eq: 4.2.4}
{\cal C}(\Delta r)\sim \left(\frac{\Delta r}{\eta}\right)^{D_2-3}
\end{equation}
The existence of a fractal measure is implied by
very general arguments from dynamical systems theory~\cite{Gra+84,Ott02,Som+93,Bec03}.
The difference between the dimension of the attractor, $D_2$, and the 
dimension of space, $d=3$, is sometimes termed the dimension deficit.
This quantity has been investigated numerically, and found to be a function
of the Stokes number, reaching a maximum value of approximately 
$3-D_2\approx 0.7$ at ${\rm St}\approx 0.7$ \cite{Bec+07}. 
The effect of particle clustering 
is argued  
to enhance the particle collision rate (\ref{eq: 4.1.6}) by a factor $ {\cal C}(2a)$
(which is large compared to unity when $a\ll \eta$, see 
Section~\ref{sec: 8.2} for more precise estimates). 
As well as numerical investigations, it is also possible to 
apply analytical techniques to calculate fractal dimensions
\cite{Wil+07,Wil+10}.

We mention here that alternative methods have been proposed to
characterize clustering~\cite{KostShaw:2001,Monchaux+10}.

\subsection{Caustics}
\label{sec: 4.3}

When inertial effects are more significant, particles can move independently of 
the fluid, and thus collide with a large relative velocity with
other particles.
As we explain here, this implies the existence of 
singularities in the 
phase-space of the suspended particles which are analogous to 
\lq caustics' in optics. These caustics can lead to a pronounced 
increase in the collision rate as the Stokes number  increases. 
It was realized~\cite{FFS02} that when inertia is large enough, 
particles ejected from vortices acquire velocities which can be 
very different from the fluid velocity, hence run into other particles with 
large relative velocity $\Delta v$.  This was termed the \lq sling effect' \cite{FFS02}. 
Later the phenomenon was described in geometric terms, using the
known notion of caustics~\cite{WM05}, which provides a convenient framework
for understanding and generalisation
(caustic structures had been 
observed earlier in models of particle suspension \cite{Cri+92}). 
Another way to discuss the phenomenon has been proposed  
in~\cite{Ijzermans09,Ijzermans10,Gus+12}. 

\begin{figure}
\includegraphics[width=0.48\textwidth]{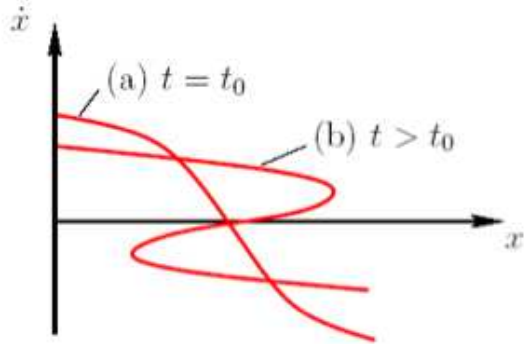}
\caption{
\label{fig: 4.2}
Phase space for a one-dimensional model: (a) The particles 
are initially on a manifold with a single-valued velocity (b) If particle
inertia is significant, particle positions may cross, because fast particles
overtake slower ones. The singularities where $\frac{{\rm d}\dot x}{{\rm d}x}$ 
diverges are termed caustics.
Creation of caustics is associated with the velocity becoming multi-valued. 
}
\end{figure}

In order to explain the role of caustics, consider a one-dimensional 
model, with a fluid velocity $u(x,t)$, and where 
virtual 
particles, which are free to pass through each other,
evolve with velocity $v$ according to (\ref{eq: 2.1}).
Consider a cloud of 
particles which are initially on a manifold in this phase space 
(see figure \ref{fig: 4.2}(a)). In the case of particles of 
negligible inertia, the velocity $v$ of particles
remains in the neighborhood of ${u}(x, t)$, so the velocity
difference between two closely separated points remains very small. 
If the particle inertia is significant, however, the velocity of particles
can become multivalued (different values of $v$ at the same
position) because faster moving particles can overtake
slower ones. This is illustrated in figure \ref{fig: 4.2}(b). 
The points where the projection of the phase space-manifold onto 
the $x$-axis is singular are the caustic singularities. 
The caustics are focal points where particles with different velocities are 
brought together at a single point. They are completely analogous to optical 
caustics, where light is partially focussed onto a line. 
Note that caustic singularities are created in pairs, and that 
between them, the velocity 
is triple-valued. The existence of finite velocity differences between particles
at the same location facilitates collisions.

In two dimensions the 
caustic points are replaced by caustic lines, where the density of particles 
diverges, illustrated in figure \ref{fig: 1.2}.
In three dimensions the caustics are sheets. 
In addition, higher-order singularities of the 
phase-space manifold, classified using catastrophe 
theory \cite{Sau80}, can also form.
For example, 
pairs of caustics can originate from a cusp singularity.

Caustics can increase the collision rate in two different ways.
Firstly,  the particle density becomes singular in the vicinity of 
the caustics at $ x_0$. If $x_0$ is the position of the caustic, the density has a 
generic divergence of the form $n(x)\sim (x-x_0)^{-1/2}$ as the caustic is approached 
from one side (here $x$ is any generic coordinate).
Secondly,  as already explained, the multivalued character of velocity can greatly
enhance the collision rate. 
The same mechanism has also been discussed in terms of ejection from 
vortices~\cite{FFS02}, and termed sling effect.
The relations between caustics and collisions of inertial 
particles were proposed in \cite{WM05,WMB06}.

The nature of the singularity at caustics can be illuminated
by considering the gradient of the velocity for nearby particles.
Assuming a continuum fluid description for the flow of particles, 
and differentiating Eq.~(\ref{eq: 2.1}) 
leads to 
the following evolution equation for the tensor $\mathbf{\sigma}$, defined
by $\sigma_{ij} \equiv \partial_i v_j$:
\begin{equation}
\frac{d \mathbf{\sigma}}{d t} + \mathbf{ \sigma }^2 = \frac{1}{\tau_{\rm p}} (\mathbf{A} - \mathbf{\sigma})
\ .
\label{eq:Burgers}
\end{equation}
where $\mathbf{A}$ is the velocity gradient tensor. Equation (\ref{eq:Burgers}),  
whose nonlinearity is of the Burgers type, is a convenient starting point 
to analyze
caustic formation numerically~\cite{FP07} or even experimentally~\cite{Bew+13}. 

To estimate the contribution of the caustic mechanism to the 
collision rate, Eq.~(\ref{eq: 2.2.1b}), we approximate the typical
velocity difference associated with the sling effect 
as $\sim \eta/\tau_{\rm K}$, 
up to a dimensionless function of ${\rm St}$ and the Reynolds 
number of the flow, ${\rm Re}$: 
\begin{equation}
\label{eq: 4.3.1}
R=\frac{4\pi a^2  n \eta}{\tau_{\rm K}} \,F({\rm St},{\rm Re})
\ .
\end{equation}
The function $F({\rm St},{\rm Re})$ must vanish in the limit as ${\rm St}\to 0$,
because the caustic mechanism is absent in the advective limit. In the 
Appendix it will be argued
that ${\rm St}\to 0$ is a singular limit, and that
\begin{equation}
\label{eq: 4.3.2}
F({\rm St},{\rm Re})\propto \exp(-S/{\rm St})
\end{equation}
provides a qualitatively plausible representation, in agreement with
numerical estimates~\cite{FP07,DP09}.

It will be demonstrated later, see Section~\ref{sec: 8}, that the
caustic or sling effect provides the dominant contribution to the collision rate,
as soon as the inertia becomes significant (for ${\rm St} \gtrsim 0.5$).

\subsection{Random uncorrelated motion}
\label{sec: 4.4}

In the limiting case where the turbulence intensity is very high,
an alternative approach to understanding the effect of increasing the
turbulence intensity was initiated by Abrahamson~\cite{Abr75}. When the damping timescale 
$\tau_{\rm p}$ is large, particles travel for long distances
through the fluid, without being influenced by the smaller scale 
turbulent eddies. Abrahamson argued out that this implies that the
particles which reach any one point will have approximately 
isotropic random velocities, so that the gas-kinetic approach 
described in section \ref{sec: 2.2.1} can be 
used to model the motion of the suspended particles.
This point has also been emphasised in~\cite{Ijzermans10}, where this regime 
was termed \lq random uncorrelated motion'.

In a turbulent flow there is a largest timescale,
$\tau_L$ (the integral timescale) which is characteristic of the driving
motion which creates the turbulence. Whenever $\tau_{\rm p}\ll \tau_L$, the 
particles will be advected with the largest eddies, and not be
sensitive
to eddies with a timescale smaller than $\tau_{\rm p}$. 
The velocity difference $\langle | \Delta v | \rangle$
in Eq.~\ref{eq: 2.2.1b} can be estimated as
the typical relative velocity of particles 
at any given point. This is comparable with the relative velocity of the 
particles and the fluid. We now describe a simple argument which can be used 
to surmise an expression for the mean magnitude of the relative 
velocity, $\langle | \Delta v |\rangle$.

We argue, in the spirit of Kolmogorov phenomenological description,
that in the inertial range the only 
parameter which enters the description of the flow is the rate of 
dissipation 
per unit mass, $\epsilon$,  together with the time scale characteristic of
the particle motion, $\tau_{\rm p}$. Dimensional arguments then 
lead to the following expression for the relative velocity:
\begin{equation}
\label{eq: 4.4.1}
\langle | \Delta v | \rangle =  {\cal K} \sqrt{ \epsilon \tau_{\rm p}  }
\end{equation}
where  ${\cal K}$ is a dimensionless constant (this approach 
was developed in \cite{Mehlig:07}).
This leads to a formula for the rate of collision in a highly 
turbulent flow:
\begin{equation}
\label{eq: 4.4.2}
R \approx K \frac{n a^2\eta}{\tau_{\rm_K}} \sqrt{ {\rm St} }
\end{equation}
The constant $K$, which is potentially important for
astrophysical applications, will be discussed in Section~\ref{sec: 8.2}. 

\section{Synthesis: a unified theory for the collision rate}
\label{sec: 8}
\subsection{Modelling the collision rate}
\label{sec: 8.1}

We have discussed several mechanisms for the role of turbulence in 
facilitating collisions between suspended particles. 
The shear component of turbulence causes collisions 
between particles moving with the fluid, at the rate which 
estimated by Saffman and Turner.
In addition, the inertial effects, parametrized by the 
Stokes number,  induce two distinct
effects.
Firstly, there is the clustering effect known as preferential concentration.
Secondly, particles located at the same spatial location may have very
different velocities, a consequence of the caustic folds in phase space. 
Finally, in the limiting case when the inertia is very large,
it was argued that 
the motions of the particles become completely uncorrelated from 
one another, and 
a simple asymptotic form for the collision rate was proposed.
In this section, we show how  to combine these competing 
mechanisms. 
The main result of the section is a derivation of a 
simple expression for the collision rate. It involves some parameters,
which must be 
determined empirically by comparison with simulations. 
We also 
consider the evidence supporting our expression coming from 
numerical simulation as reported in \cite{Vosskuhle+14a}. 

The central idea is that the collision rate can be resolved into two
components. Some collisions arise from particles which follow similar 
trajectories for an extended period, and which eventually come into 
contact because of shearing motion in the flow. At low Stokes number,
where particles are exactly following by the flow, this is the 
only mechanism.
The collision rate due to this mechanism is denoted $R_{\rm adv}$, 
and it is controlled by the local shear rate, as well as by the local
concentration around particles.
Collisions between particles not following fluid path lines, 
occurring when caustics start to form in the phase-space of 
the suspended particles,  give rise to
a very different contribution, denoted here by $R_{\rm caust}$.
The two contributions $R_{\rm adv}$ and $R_{\rm caust}$ differ in an essential 
way by their dependence on the siza $a$ of particles.
We argue that these mechanisms operate independently and that their
contributions are additive, so that
\begin{equation}
\label{eq: 8.1}
R=R_{\rm adv}+ R_{\rm caust}
\ .
\end{equation}
This decomposition was proposed by 
analysing theoretical models~\cite{WMB06,Gustavsson:11} 
and simulations of simplified numerical models~\cite{DP09}.
There is no hard criterion which distinguishes  between 
an advective and a caustic-mediated collision, so (\ref{eq: 8.1}) must be 
regarded as an approximation. 
In the following, we derive the form of the terms $R_{\rm adv}$ 
and $R_{\rm caust}$ in (\ref{eq: 8.1}).

In the limit ${\rm St} \ll 1$ the collision rate is determined by shearing
motion, without any inertial effect, and the collision rate is given
by equations (\ref{eq: 4.1.6})~\cite{ST56}.

The preferential concentration causes clustering of particles with finite
values  of ${\rm St}$. The density of particles
at a distance $r$ from a given test particle is  $n {\cal C}(r)$, where
${\cal C}(r)$ is a radial correlation function.
An important remark is that the inertial effect that leads to
clustering 
does not enhance significantly the relative velocity of particles, contrary 
to caustics. In fact, particles with large velocity
differences do not stay together, thus not contributing to 
preferential concentration.
As a result, the enhancement of the local concentration around
particles contributes only the $R_{\rm adv}$ term, which becomes:
\begin{equation}
\label{eq: 8.3}
R_{\rm adv}= \sqrt{ \frac{8 \pi}{15} }\frac{ n (2 a)^3}{\tau_{\rm K}}  {\cal C}(2a)
\ .
\end{equation}
When inertial particles converge to a fractal measure, as we 
have argued, see Section.~\ref{sec: 4.2}), the function $ {\cal C}(r)$ behaves as 
a function of $r$ as a power-law:
${\cal C}(r) \propto r^{-D_2-d}$.

In the caustic-dominated case we 
combine expressions (\ref{eq: 4.3.1}), (\ref{eq: 4.3.2}) and (\ref{eq: 4.4.2})
to obtain an expression for the rate of collisions due to caustics:
\begin{equation}
\label{eq: 8.4}
R_{\rm caust}=K\frac{ n a^2\eta}{\tau_{\rm K}}\sqrt{{\rm St}}\exp(-S/{\rm St})
\end{equation}
where $K$ and $S$ are two constants to be determined by fitting
the collision rate against numerical simulations: $K$ determines the 
asymptotic collision rate at very large Stokes numbers, and $S$ determines
the crossover point where the caustic mechanism becomes significant.
The non-analytic term is consistent with theoretical expectations \cite{WMB06} 
and numerical studies \cite{FP07}.

Taken together, equations (\ref{eq: 8.1})-(\ref{eq: 8.4}) define the theory for the 
collision rate.  We note that the function $F({\rm St}, {\rm Re})$ 
defined in equation (\ref{eq: 4.3.1})
is known only through its asymptotic behaviour in certain regimes, so 
further information is necessary to make progress.  
The two terms in (\ref{eq: 8.1}) differ in an essential way through
their dependence on the size of the particles $a$, at a fixed value of
${\rm St}$ and ${\rm Re}$. Namely, $R_{\rm caust}$ varies as $a^2$, wheres
$R_{\rm adv}$ varies as $a^{D_2}$ ($D_2 > 2$). This difference can be 
used to separate the different contributions numerically.
We present in the following section numerical results, which indicates 
that the decomposition (\ref{eq: 8.1}) is indeed a very effective tool for the 
analysis of collision rates. 
We note that Zaichik and co-workers \cite{zaichik:03,zaichik:10} proposed 
models of the rate of collision in turbulent flows which 
use similar physical principles. Their final expressions for 
the collision rate involve a much larger number of parameters
than equations (\ref{eq: 8.1})-(\ref{eq: 8.4}).

As well as the collision rate, the probability distribution of relative 
velocities of colliding particles is also of interest, especially in the context
of understanding planet formation, where collisions may be sufficiently 
energetic to cause fragmentation \cite{Gus+08,Pan+13}.

\subsection{Numerical evidence on collision rates}
\label{sec: 8.2}

A substantial amount of literature has been devoted to the numerical
investigation of the collision 
rates in both direct numerical simulations (DNS) of the Navier-Stokes
equations, and model flows, but many of these studies
pre-date some of the theoretical insights contained in equations
(\ref{eq: 8.1})-(\ref{eq: 8.4}) above. 
The collision rate does show a marked increase as the effects inertia 
are increased, and this is usually ascribed to the effects of \lq preferential
concentration', that is the clustering effect, but it has been argued that effects 
of caustics may also be  significant~\cite{Sundaram:97}. 
Here we summarise some recent numerical studies which separate out
the two contributions in equation (\ref{eq: 8.1}) by studying the dependence of the 
collision rate
upon particle size, keeping the Stokes number fixed. 
This is achieved by varying the
ratio of the particle density to the fluid density, $\rho_{\rm p}/\rho_{\rm f}$. 
Modifying the ratio $\rho_{\rm p}/\rho_{\rm f}$ at fixed value of
the Stokes number is achieved by varying
in the collision detection algorithm the radius of the
particles, $a$, according
to (\ref{eq: 2.2}),(\ref{eq: 2.8}) (so that
$a\propto (\rho_{\rm p}/\rho_{\rm f})^{-1/2}$).

\begin{figure}
\begin{center}
\includegraphics[width=0.80\textwidth]{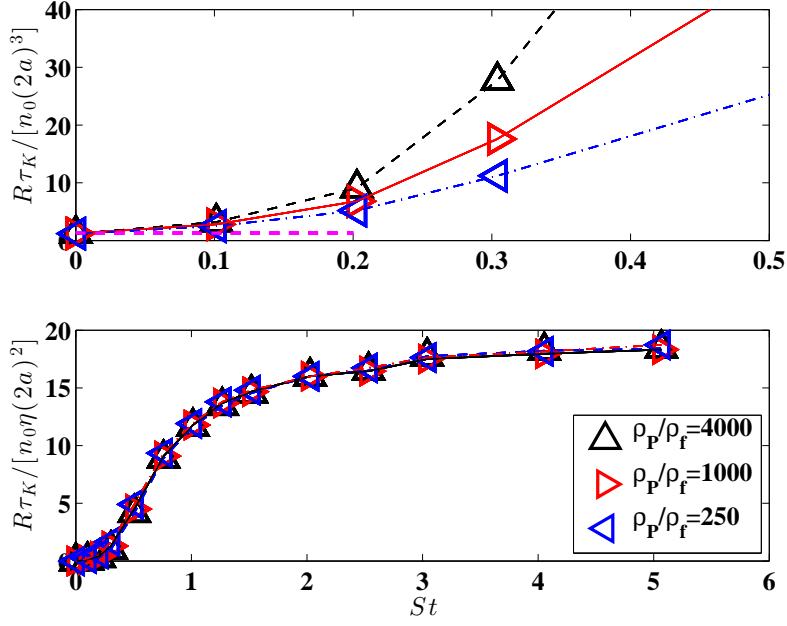}
\caption{\label{fig: 8.1}The collision rate $R$ as a function of the Stokes number
${\rm St}$, for three different values of the ratios of density
$\rho_{\rm p}/\rho_{\rm f} = 250$, $10^3$ and $4\times 10^3$.
The collision rate $R$ is normalized by
$ n (2a)^3/\tau_{\rm K}$ (a), and
$ n  (2a)^2 \eta/\tau_{\rm K}$ (b).
The horizontal dashed line in (a) corresponds to the Saffman-Turner prediction.}
\end{center}
\end{figure}

The collision rate, $R$, is determined numerically by recording among
a set of trajectories, all instances in which the separation radius decreases 
past $2a$. In the case of collisions where particles stick or coalesce on 
contact, we should only count the first contact collisions. This effect 
should be accounted for by introducing a factor $f<1$ in (\ref{eq: 8.1}). 
The coefficient is no smaller than $\approx 0.85$ when ${\rm St}$ 
is very small, and decreases when ${\rm St}$ increases~\cite{Vos+13}.
Here, we do not distinguish between single and multiple collisions. 
The collision rate, $R$, shown in Fig.~\ref{fig: 8.1}(a), is normalized by
$ n (2 a)^3/\tau_{\rm K}$ and plotted as a function of ${\rm St}$.
The Saffman-Turner prediction, (\ref{eq: 4.1.6}), implies that in the limit 
${\rm St}\rightarrow 0$, the quantity $R \tau_{\rm K}/(n (2a)^3)$ should become
independent
of the ratio $\rho_{\rm p}/\rho_{\rm f}$. Our own numerical results 
\cite{Vosskuhle+14a}, in agreement with other 
estimates~\cite{Sundaram:97,Wang:00},  are
only consistent with this prediction for small values of ${\rm St}$.
Fig.~\ref{fig: 8.1}(b) shows
that $R \tau_{\rm K}/(n a^2\eta)$
as a function of the Stokes number,
does not depend much on $\rho_{\rm p}/\rho_{\rm f}$ for values of ${\rm St}$
larger than $\gtrsim 0.3$.
This scaling is consistent with the sling/caustics
collision mechanism, described by equation (\ref{eq: 8.3}).
We note that $F({\rm St},{\rm  Re})$ deduced from Fig.~\ref{fig: 8.1}(b) does not
fit the asymptotic
form $F({\rm St},\infty)=K\sqrt{{\rm St}}$ for large values of ${\rm St}$. We ascribe
this to the limited Reynolds number of our numerical simulations.

\begin{figure}
\includegraphics[width=0.80\textwidth]{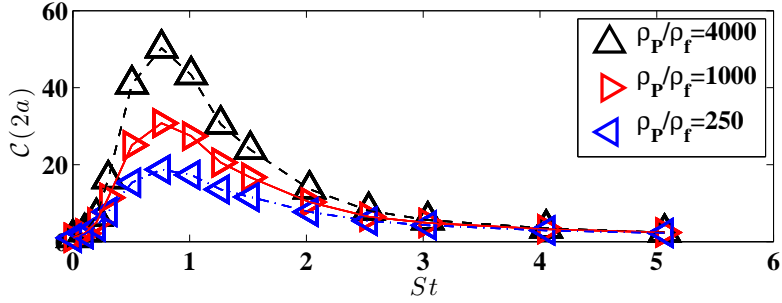}
\caption{The function ${\cal C}(2a)$ that measures preferential concentration,
for three values of the density ratio, $\rho_{\rm f}/\rho_{\rm f}$, as indicated in the figure.
The preferential concentration does not play a significant role for
${\rm St} \gtrsim 5$
\label{fig: 8.2}
}
\end{figure}

Figure \ref{fig: 8.2} shows the effect of clustering in the same simulations: 
the effect of clustering on the function ${\cal C}(2a)$ reaches a maximum 
at a value ${\rm St} \approx  0.7$ of the Stokes number. The
collision rate grows much larger at higher values of the Stokes number,
which provides further indication 
that caustic effects dominate the collision rate.

\begin{figure}
\includegraphics[width=0.80\textwidth]{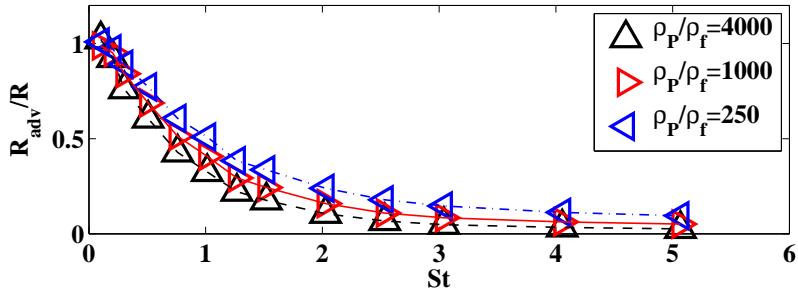}
\caption{The ratio of the contribution to the collision
rate due
to preferential concentration $R_{\rm adv}$, defined by 
equation~(\ref{eq: 8.3})
and of the total collision rate, $R$. At fixed value of
$\rho_{\rm p}/\rho_{\rm f}$,
The contribution of $R_{\rm adv}$ to the total collision rate decreases
when the Stokes number increases, or when the ratio $\rho_{\rm p}/\rho_{\rm f}$
increases.
\label{fig: 8.3}
}
\end{figure}

Figure \ref{fig: 8.3} shows the relative importance of the advective and 
caustic terms in these simulations. The caustic contribution becomes dominant 
for Stokes numbers greater than $\approx 0.75$.  

At very large Stokes numbers the collision rate is expected to be given by 
(\ref{eq: 4.4.2}), which is a potentially important prediction
for astrophysical applications. 
To estimate the unknown constant $K$ in (\ref{eq: 4.4.2}), one needs 
simulations at high enough ${\rm St}$, i.e., high enough $\tau_p$, but with 
the constraint that the value of $|\mbox{\boldmath$R$}|$, such that 
$\tau_p = (|\mbox{\boldmath$R$}|^2/\epsilon)^{1/3}$,
is in the inertial range. Extrapolation of numerical data at
moderate Reynolds numbers from different
sources~\cite{Vosskuhle+15} leads to a value $K \approx 50$.

An alternative decomposition of the collision rate, originally proposed by~\cite{Sundaram:97},
expresses
the collision rate $R$ as a product in which the term ${\cal C}(2a)$, which
describes the
local concentration enhancement around a particle, appears as
an overall factor:
\begin{equation}
\label{eq: 8.7}
R =  2 \pi (2a)^2  n {\cal C}(2a) \langle  | \Delta v | \rangle_{\rm eff}
\end{equation}
This representation, which is exact for a suitable definition
of $\langle  | \Delta v | \rangle_{\rm eff}$, suggests that
the preferential concentration and sling effects act together to
enhance the collision rate. Fig.~\ref{fig: 8.1}(b)
demonstrates that if this parametrisation of the collision
rate is used, then the dependence of ${\cal C}(2a)$ upon $\rho_{\rm p}/\rho_{\rm f}$,
see Fig.~\ref{fig: 8.2} must be cancelled
(for ${\rm St} \gtrsim 0.5$) by a reciprocal dependence of the collision 
velocity, $\langle  | \Delta v | \rangle_{\rm eff}$.
In fact,
previous measurements~\cite{Bec:10,Rosa:13} of the dependence of ${\cal C}(r)$
and of the average velocity difference as a function of $r$ suggest power law
dependences, the exponents being such that the product
${\cal C}(2a) \langle  | \Delta v | \rangle_{\rm eff}$
is essentially constant for ${\rm St } \gtrsim 0.5$. 
Equations
(\ref{eq: 8.1}), (\ref{eq: 8.3}) and (\ref{eq: 8.4})  provide
a very natural explanation of this cancellation. 
We remark that the power-law dependence
of the collision velocity has been explained in a random flow
model~\cite{Gustavsson:11,Gus+13}, and used to justify Eq. (\ref{eq: 8.1})
for that system. 

We  therefore conclude that the 
decomposition (\ref{eq: 8.1}), which rests on a physically well-motivated
analysis, and which is supported by the analysis of simplified theoretical 
models, provides a consistent description of the available numerical data. 
One of the main lessons from the analysis of the dependence of the 
collision rate on ${\rm St}$ and on the ratio $\rho_{\rm p}/\rho_{\rm f}$, is that 
the sling/caustic effect provides the dominant mechanism for the dramatically
enhanced collision rate of particles whose Stokes number exceeds 
$\gtrsim 0.75$, in the cases of water droplets.

\section{From collisions to aggregation}
\label{sec: 9}

Up to this point we have considered expressions for the rate of geometrical
collisions. In order to determine the aggregation of particles, we need to 
consider whether the collisions result in particles combining, and how 
the population of particle sizes evolves.

\subsection{Collision efficiency }
\label{sec: 9.1}

Droplets which undergo a geometrical collision, {\it i.e.} whose 
separation falls below $a_1+a_2$ when their motion is predicted 
by equation (\ref{eq: 2.1}),
might not coalesce, because the streamlines of 
small droplets curve around larger ones. In fact, if the Navier-Stokes 
equations were a complete description, droplets would never
collide, because 
 of the presence of a lubricating film of fluid between them. 
Other mechanisms allow the particles to actually coalesce~\cite{Chun+05}.

The aggregation of particles is characterised by a {\em collision 
efficiency}  $e$, defined as the ratio of the observed rate 
of coalescence 
to the rate of geometrical collisions (where the separation of the centres 
predicted by Eq. (\ref{eq: 2.1}) falls below $a_1+a_2$). The coalescence efficiencies are hard to determine, 
because the the complex physics determining breakdown of the lubricating 
layer. It is widely accepted that they are low for typical cloud 
droplets \cite{Mas57,Pru+97}.
If the larger droplet has radius below $20\,\mu{\rm m}$, it is believed that $e \le 0.1$,
and that for radius $10\,\mu {\rm m}$, $ e \le 0.03$ \cite{Pru+97}. 
For droplets of size
$a=50\,\mu {\rm m}$ colliding with droplets of size $a=10\,\mu {\rm m}$, 
however,
the efficiencies are expected to be close to unity \cite{Mas57,Pru+97}.

\subsection{Models of aggregation}
\label{sec: 9.2}

The theory developed by Smoluchowski~\cite{Smol:17} to
describe the coagulation in colloids seems to provide a general framework
to discuss a vast class of problem, including coalescence of water 
\cite{Shaw03,Derevyanko08}
or of dust grains. The approach of~\cite{Smol:17} is naturally formulated
in terms of mass conservation, so we  use here $\bar N(m,t)$ the
number of particles of mass $m$ per unit volume at instant $t$. The
Smoluchowski equation is:
\begin{eqnarray}
\label{eq: 9.1}
\frac{\partial  \bar N}{\partial t}(m,t) & = & 
\frac{1}{2}\int_0^m {\rm d}m'\ K(m-m',m')  \bar N(m',t)  \bar N(m-m',t) \\ \nonumber
& - &  \bar N(m,t)\int_0^\infty {\rm d}m'\ K(m,m')\bar N(m',t)
\ .
\end{eqnarray}
The Smoluchowski description is a mean-field model, in that it 
assumes that the
particle density is spatially uniform. To treat aggregation processes,
the Smoluchowski approach is generally preferred
to the full master equation, which is completely
intractable.

Depending on the structure of the collision kernel $K(m.m')$, the Smoluchowski
equation can predict one of two different types of particle growth.
In {\em ripening} processes, the particles grow but they remain of comparable 
sizes. Alternatively, if the collision kernel $K(m,m')$ increases sufficiently 
rapidly as $m, m'\to \infty$,
the growth process is unstable: a small fraction of particles undergoes runaway
growth in a finite time. This means that, in the Smoluchowski model, all
 of the smaller particles end up absorbed into one
massive cluster in a finite time. 
This phenomenon, known as {\em gelation}, has been investigated intensively: 
see \cite{K+10} 
for a review. In the notation of Eq.~(\ref{eq: 9.1}), gelation is expected
to happen when the collision rate $K(m,m')$ grows faster with $m$ and $m'$
than linearly~\cite{Aldous99}. This is of primary concern in our case, since
for large values of the mass, or equivalently, of the radius, the 
dominant physical effect in the collision rate is due to particle settling, 
see Eq.~(\ref{eq: 2.5}).  In this case, the kernel $K$ grows as
$K(M, m) \propto M^{4/3}$ when $M$ is large, so the mean-field
equation (\ref{eq: 9.1}) leads to gelation.
Here we discuss the consequences of the gelation transition in the Smoluchowski
model. 

The requirement of a very large number of collisions, of the order of 
$10^6$, to form a raindrop of typical size $1\,{\rm mm}$ out of much smaller
microscopic droplets (of typical size $10\, \mu{\rm m}$) makes gelation
an appealing feature of the Smoluchowski equations.

There is, however, a potential problem with using the Smoluchowski 
equation to model rainfall. The gelling transition occurs after a critical 
time $t_{\rm c}$. Investigations of the gelation transition for the 
Smoluchowski equation for {\em homogeneous kernels} of the form
\begin{equation}
\label{eq:9.3}
K(m_1,m_2)={\rm const.}\ m_1^\mu m_2^\nu
\end{equation}
have shown rigorously~\cite{vDo87,Carr92} that there is a gelation 
transition when 
$\mu+\nu>1$, and that $t_{\rm c}=0$ (termed {\em instantaneous gelation})
whenever $\nu>1$ or $\mu >1$. The kernel for liquid droplets 
settling under gravity is asymptotic (when $m_1\gg m_2$) to a homogeneous
 kernel with $\nu=4/3$, so the issue of instantaneous gelation is relevant.

These results suggest that the Smoluchowski
equations, which predict an {\em instantaneous} gelation, must to be
used with caution.

We should consider the reason why the Smoluchowski 
equation has this pathological behaviour when $\nu>1$. 
Two observations
are necessary to explain the origin of the zero-time singularity. First, 
note that a cluster of size $M$ will grow to infinite size in a time  
$t_{\rm c}(M)$, which should be a decreasing function of $M$ 
(because the rate of collisions is assumed to increase with the cluster size).
Furthermore, under the conditions where instantaneous gelation occurs, this 
time $t_{\rm c}(M)$ approaches zero as $M\to\infty$. 
The second observation
is that, in an infinite system, a cluster of size $M$ arises in a 
finite time $t_0$ somewhere in the system, no matter how small we 
choose $t_0$: we can find a cluster 
of size $M$ somewhere in an infinite system, no matter how small we make 
$t_0$ or how large we make $M$. According to the Smoluchowski equation, the
model thus undergoes gelation at a time 
$t=t_0+t_{\rm c}(M)$ where both $t_0$ and $t_{\rm c}$ can be made 
arbitrarily small. This argument applies to a system with
an infinite number of particles: see \cite{Bal+11} for a discussion
of finite-size effects.

We note that the prediction of a runaway growth may need to be
revised, because when the 
droplets become very large, the terminal velocities are proportional to
$a^{1/2}$, so the collision kernel grows as $m^{5/6}$, which does not 
yield a gelation transition. In addition, rain droplets 
may undergo fragmentation when they become sufficiently large~\cite{Villermaux:09}.

This pathology of zero-time gelation results from the use of a 
mean-field description, which ignores any 
spatial structure of the particle density $ {\bar N}(m,t)$. Under 
the mean-field approximation, the local depletion of particles due to the
formation of a large cluster is not taken into account, so an unbounded growth
can proceed. We conclude that the Smoluchowski equation 
must be used caution, and that a proper description of the runaway growth therefore requires
more realistic assumptions about the spatial distribution.

\section{Application to rain initiation}
\label{sec: 10}

The problem we are focusing on here concerns the generation of rain drops
in \lq warm' (ice-free) cumulus clouds \cite{Mas57,Pru+97,Rog+82,Shaw03}. 
In ice-bearing clouds the Wegener-Bergeron-Findeisen can result in rapid 
growth of ice crystals by condensation, which makes it easier to explain 
precipitation~\cite{Mas57}.
When the air becomes supersaturated, water droplets condense
rapidly onto aerosol nuclei. They grow up to a typical radius 
$10\,\mu{\rm m}$ before the supply of water vapour is exhausted.
The later stages of the growth of a raindrop (from radius of approximately 
$50\,\mu{\rm m}$ upwards) involve the falling drop growing rapidly by coalescence with 
microscopic droplets lying in its path, and are easier to understand. 
The challenge is to explain the growth of droplets from
size $10\,\mu{\rm m}$ to approximately $50\,\mu {\rm m}$, and in particular,
to identify the physical process leading to the rapid onset of rain showers,
which can develop in less than half an hour.

To illustrate the discussion, consider the following representative values
for a convecting cumulus cloud which could produce precipitation 
\cite{Mas57,Shaw03}. The typical
droplet radius is $a=10\,\mu {\rm m}$, the number density of droplets is 
$n=4\times 10^8\,{\rm m}^{-3}$, and the cloud depth is $L=10^3\, {\rm m}$.
The typical vertical velocity of air inside the cloud has magnitude
$2\,{\rm m}\,{\rm s}^{-1}$, so that the eddy turnover time may be taken to be
$\tau_L=10^3\,{\rm s}$. An estimate for the rate of dissipation is
$\epsilon\approx L^2/\tau_L^3=10^{-3}\,{\rm m}^2{\rm s}^{-3}$,
which gives an estimate of the Kolmogorov time 
$\tau_{\rm K}\approx 10^{-1}\,{\rm s}$.
Rain falls as droplets of size approximately $1\,{\rm mm}$. 

What are the processes possibly leading to the growth of very small 
droplets with an initial size of typically $10\,\mu{\rm m}$? 
Growth can proceed by collisions, and one possibility is that droplets
settling at different rates lead to collisions. This mechanism is only possible 
in the presence of size dispersion among droplets. 
Specifically, inserting values for air and water at $5^\circ {\rm C}$ 
into equation (\ref{eq: 2.4}) gives $\kappa\approx 1.4\times 10^8\,{\rm m}^{-1}{\rm s}^{-1}$.
Then, using the expression of the collision rate (\ref{eq: 2.5})
to estimate the coalescence rate of a droplet of 
radius $a+\Delta a$ (with $a = 10\,\mu {\rm m}$, $\Delta a \approx 2.5\, \mu{\rm m}$) 
falling through a gas of smaller droplets (of radius $a = 10\,\mu {\rm m}$),
and with a collision efficiency
$e \approx 0.03$, gives a rate of coalescence  
$R \approx 10^{-4}{\rm s}^{-1}$.
We conclude that the rate of coalescence
of typical sized water droplets induced by differential settling is 
very small, as long as the dispersion between particles is small.

As the air in a convecting (cumulus) cloud is turbulent, it may be expected 
that the coalescence
rate between droplets may be greatly facilitated by turbulence~\cite{ST56}.
Using the expression for the collision rate between small droplets 
following the flow, (\ref{eq: 4.1.6}), and
the parameters of the cloud model, gives 
$R_{\rm turb}\approx 2\times 10^{-6}\,{\rm s}^{-1}$,
which is negligible. The effects of turbulence are dramatically increased
when the effects of droplet inertia are significant. Inertial effects are measured
by the Stokes number, ${\rm St}\equiv\tau_{\rm p}/\tau_{\rm K}$. 
The collision rate is greatly enhanced by effects due to caustics for 
${\rm St} \gtrsim 0.3$ \cite{Vosskuhle+14a}. With the values of cloud model
given above, however, the estimated Stokes number is 
${\rm St}\approx 10^{-2}$, which is to small to lead to any significant 
enhancement.

The small rate of collisional coalescence of droplets is clearly a problem.
It should be kept in mind that considerable variation exists among 
various clouds, and even within a given cloud, the flow is expected
to be very inhomogeneous. The numbers used in the estimates above could
therefore vary significantly, resulting in a large 
increase of the coagulation rates, compared to the values given here, 
at least in parts of the clouds. One may also remark that 
only a very small proportion
of the microscopic droplets needs to be converted into raindrops.
Consider the rate at which droplets actually form.
Rainfall at a rate of 
$3.6\,{\rm mm}\,{\rm hr}^{-1}=10^{-6}\,{\rm m}\,{\rm s}^{-1}$
is considered as \lq moderate' \cite{Mas57}. If the raindrops have size 
$a\approx 1\,{\rm mm}$,  the number of drops falling
per second and per square meter is approximately $250$.
Given the assumed cloud depth of $L=10^3\,{\rm m}$, the volumetric
rate of production of raindrops is approximately 
$0.25\, {\rm m}^{-3}{\rm s}^{-1}$. If the
microscopic droplets have density $ n=4\times 10^8\,{\rm m}^{-3}$, then the rate of conversion
of each microscopic droplet into a \lq collector' droplet undergoing runaway 
growth is
approximately $6\times 10^{-10}\,{\rm s}^{-1}$.
Alternatively, during a five minute shower, the 
probability that a water droplet starts growing, and accumulates during its subsequent
fall enough droplets to become a rain droplet is small,
approximately $2\times 10^{-7}$.
The problem of rain initiation is, therefore,
concerned with the frequency of very rare events~\cite{Kos+05}.

Despite the
fact that the required conversion probability is very small (of order $10^{-7}$),
growth of droplets is too slow by a collisional mechanism.
On growing from $10\,\mu {\rm m}$
to $50\,\mu{\rm m}$, the volume of a droplet increases by a factor of $125$,
that is, there are or order $100$ collision events.
It was argued above that the rate for the first collision events is small,
$R\approx 10^{-4}\,{\rm s}^{-1}$. It is not obvious whether the rarity 
of the event is sufficient to compensate for the low rate of multiple 
collisions. Large deviation theory (reviewed in \cite{Tou09}) is the appropriate 
tool for analysing this problem. A large deviation analysis shows that a sufficient 
number of droplets can undergo runaway growth in small fraction of the 
mean time to the first collision~\cite{Wil15a}.

After a droplet has grown to a size much larger than the typical 
droplet size, it falls rapidly and collects other droplets
in its path. When $a>50\,\mu{\rm m}$ we assume that the collision efficiency is
approximately unity.  
Consider a droplet of size $a_1$ falling through a \lq gas' of small droplets,
which can be characterised by the liquid volume fraction $Q=4\pi  n\langle a^3\rangle/3$.
The large droplet falls with velocity $v=\kappa a_1^2$ and grows in volume at
a rate $\pi a_1^2 Q v$, so that
\begin{equation}
\label{eq: 10.4}
\frac{{\rm d}a_1}{{\rm d}t}=\frac{\kappa Qa_1^2}{4}
\ .
\end{equation}
Solving this equation shows that the droplet radius diverges in the time
\begin{equation}
\label{eq: 10.5}
\tau_{\rm exp}=\frac{4}{\kappa  Q a_1}
\ .
\end{equation}
Equation (\ref{eq: 10.5}) predicts that the time before runaway increases 
rapidly as the droplet size decreases. For 
the parameters introduced to describe a cloud, 
a droplet of size $a_1=50\,\mu {\rm m}$
requires $\tau_{\rm exp}\approx 2 \times 10^3\,{\rm s}$ to undergo 
explosive growth.

We conclude that, although runaway growth can proceed
when a droplet reaches $50\,\mu{\rm m}$, it is difficult to understand how
droplets can reach this size in the short time that it takes for a rain shower
to develop. While turbulence mediated  collisions may be 
an important ingredient to explain the runaway growth of rain drops in 
warm cumulus clouds, a complete picture is likely to involve other effects,
possibly including those triggered by 
non-collisional growth processes \cite{Wil15}. Such effects need to be
discussed more systematically in the future.

\section{Application to planet formation}
\label{sec: 11}

The other major area for applications of turbulent collision processes
is in understanding planet formation. Here we can only give 
a brief introduction to a complex and rapidly developing research 
field. We start by summarising the standard 
model, which is reviewed in~\cite{Arm10}, before discussing the unresolved difficulties.

When a star forms by gravitational collapse, conservation
of angular momentum prevents all of the cloud of gas from falling into the star,
and the residual material rapidly forms a disc-like structure (which has the 
least internal motion
consistent with the conservation of angular momentum).  
It is assumed that  planet formation occurs in such circumstellar 
discs surrounding young stars \cite{Safranov:69}, which implies
that planets 
would be created with near circular orbits in the plane of the disc.
This model
is consistent with the structure of the solar system, where the planets lie
in approximately circular and coplanar orbits, close to the equatorial plane
of the Sun. The structure of these circumstallar discs
is described by a model introduced by Shakura and Sunyaev \cite{Sha+73}.

The interstellar medium from which planets form 
is thought to contain sub-microscopic dust. The grains have a broad 
distribution of sizes,
but $10^{-7}\,{\rm m}$ is a typical value. 
The grains may be ice particles or minerals,
predominantly silicon or carbon based, originating from nuclear
reactions in stars.
The proportion of \lq heavier' elements (that is, elements 
other than ${\rm H}$ and ${\rm He}$)
in star forming regions is of the order of one percent by mass. 
There is also a consensus that the dust grains play
an important role in planet formation. 
It is assumed that dust grains adhere on contact, making ever larger structures.
In the simplest version of this model, these objects continue to accumulate material
until they become kilometre-sized \lq planetesimals', and gravitational 
forces between 
the planetesimals take over.
A variation on this model suggests that \lq boulders' settle to form a 
layer at the mid-plane of the circumstellar disc. When the density
of material at the mid-plane reaches a critical level, this triggers a 
gravitational collapse \cite{Gol+73}.

Collisions between grains to produce larger structures play an essential role 
in these routes to planet formation, and the aim is to understand 
whether a quantitative description of these processes is viable.
In addition to the difficulty in estimating the physical parameters 
inside a circumstellar with current observational techniques,
some serious theoretical difficulties with these models are hard 
to circumvent \cite{Wil+08}. 

In the absence of any turbulent transport mechanism, the
rate of collision of microscopic grains is very low. 
It seems likely, but it is not certain, that the gas in a circumstellar
disc is in turbulent motion: the Reynolds numbers associated with the motion
are extremely high, and the effects such as convection due to frictional 
heating will combine
with the rotational motion.  
Turbulence certainly increases the rate of collisions,
but it also brings its own problems \cite{Blu+08}. The dust particles will 
adhere to each other
due to van der Waals forces and to a certain extent electrostatic forces.
Because these binding forces are weak, aggregates of dust particles are 
very easily fragmented by collisions. Estimates of the relative velocity of 
particles colliding
in a turbulent environment such as equation (\ref{eq: 4.4.1}) indicate that the 
collision speed increases with the size of the particles 
\cite{Vol+80,Mehlig:07}, so that there may be a maximum value for
the size of a dust cluster which can be formed by aggregation in a turbulent
environment. Recent estimates indicate that the maximum size that can be 
reached by
clusters of dust particles appears to be very small for reasonable values of the
parameters in a model for the protoplanetary accretion disc. 
It is useful to give some estimates for conditions inside the circumstellar
disc according to the Shakura-Sunyaev model. These have a power-dependence
upon distance from the star, but at $1\,{\rm AU}=1.5\times 10^{11}\,{\rm m}$ 
({\em i.e.}, the Earth-Sun distance)
the gas density is $3\times 10^{-4}\,{\rm kg}\,{\rm m}^{-3}$, the dissipation rate 
is $10^{-3}\,{\rm m^2}\,{\rm s}^{-3}$, the gas mean-free-path is $4\times 10^{-3}\,{\rm m}$,
the speed of sound is  $700\,{\rm m}\,{\rm s}^{-1}$ \cite{Wil+08}. The low density of the
gas implies that the Epstein formula (\ref{eq: 2.3}) is applicable, and the 
particle relaxation time 
$\tau_{\rm p} $ is very large. Thus, equation 
(\ref{eq: 4.4.1}) implies that the relative velocity of colliding particles 
is quite large. For example $30\,{\rm cm}$ size objects are predicted to 
collide with velocities
of approximately $10\,{\rm m}\,{\rm s}^{-1}$ \cite{Wil+08}. It seems improbable that balls of dust
particles would survive collision at these speeds.
These considerations suggest that there are severe theoretical difficulties with models
based upon aggregation of dust particles.

Other outstanding theoretical difficulties are related to the model 
itself, and are independent of whether turbulence 
plays a role in collisional aggregation. One of the difficulties 
concerns the fact that the gas in an accretion
disc is partially supported by its pressure, so that its orbital velocity is
slightly lower than the Keplerian value in the quasi-static state. A \lq rock'
(more accurately, an aggregate of dust grains and possibly ices) which is
entrained with the gas is not supported by the pressure and consequently slowly
spirals in towards the star \cite{Wei77,Tak+02}. This effect is most pronounced
for rocks with size comparable to the mean free path of the gas
(typically $1\,{\rm cm}$ to $1\,{\rm m}$), and the timescale for spiralling in
is of the order of $300\,{\rm yr}$ starting from an orbit at $1\,{\rm AU}$.
Even under the most favourable assumptions about growth rates by aggregation, it
is difficult to see how aggregates of dust particles can grow sufficiently
rapidly to avoid spiralling in. A variety of complicated theories have been
proposed to try to resolve this difficulty. 
As of now a \lq streaming instability'~\cite{You+05, Joh+12} is regarded as a 
promising theoretical approach.

More recently, the detection of a large number of extra-solar 
planets \cite{Win+15} has brought new challenges.
These discoveries yielded many surprises, some of which are hard to reconcile 
with the standard model for planet formation. 
Significant numbers of these exoplanets have large orbital eccentricity.
Various models have been proposed to account for this \cite{Zak+04}.
The most plausible of these is a slow-acting three-body instability
resulting in a drift of orbital parameters, leading to a near-collision
between two planets. This could cause escape of one planet and scattering of
the other to an eccentric (and probably non-equatorial) orbit \cite{For+03}.
It is as yet not clear whether the large proportion of exoplanets with
eccentric orbits can be explained by this model. The model would suggest
that large planets are less likely to be scattered into highly eccentric
orbits. There seems, however, to be a positive correlation between eccentricity
and mass, which lends support to alternative mechanisms of planet formation~\cite{Rib+07}. 
There are also numerous examples of planets where the orbital
plane is at a very large angle of inclination to the equator of the star, or where the
orbit of the star is in the opposite direction to the spin direction of the star \cite{Win+15}. 
These observations cast a doubt on one of the central hypothesis
of the standard planet formation model \cite{Safranov:69}, which imply that planets are formed in 
circular orbits in the circumstellar disc.

As for rainfall, understanding the role of turbulence in mediating collisions
has not resolved the difficulties in explaining planet formation. If the dust accretion
model will ultimately be shown to be correct, collisions in turbulent flows will 
play an important role. It is possible, however, that alternative theories will
be required \cite{Rib+07,Wil+12}.

\section{Perspectives for future work}
\label{sec: 12}

The recent developments 
in the theory of collision between particles in turbulent suspensions presented 
here
correspond to quantitative progress in describing a complex physical 
phenomenon. The collision kernel for tracer particles which
are exactly following the turbulent flow had been understood for many
years. 
In contrast, the problem collisions between inertial
particles, which is much more realistic for numerous applications,
involves subtle physical effects, which can be traced back to the
fact that the trajectories deviate from the fluid ones.
The first effect, preferential concentration,  i.e. the tendency 
for particles to be unevenly 
distributed in the flow, had been noticed in many experiments and 
simulations. It can be qualitatively understood in terms of fractal 
attractors. The notion of slings or caustics, leading to strong velocity
difference between colliding particles, has emerged from a combination
of theoretical and numerical work which we have reviewed in this paper. Over
a vast range of particle inertia, the latter effect is the dominant one 
to determine the large collision rate between particles. While the 
understanding of these phenomena was largely motivated by a problem of cloud 
physics, it is very likely that the knowledge developed here 
will find other applications in other fields of science. 

While the knowledge of collision kernel is a crucial first step to understand
coagulation of particles in a turbulent suspensions, an accurate description
of the actual formation and growth of clusters rests on the solution of
kinetic models. The usual approach is a \lq mean-field' description originally
proposed by Smoluchowski. Given
the functional dependence of the collision kernel in problems inspired from
cloud physics, this model predicts
a singular behavior at zero time, which points to difficulties in the use of the model.
The runaway growth of a small minority of drops is however expected
to be an important feature in cloud physics. Identifying the mechanisms which 
trigger a runaway  growth, and quantifying how frequently this 
happens are 
significant areas for further development.

The knowledge gained from the description of the 
collision kernel in turbulent suspensions has not yet led to an unambiguous
understanding of the formation of large clusters in the
important physical contexts of rainfall and planet formation. This situation calls for future work. 
Future theoretical developments may be necessary, in particular 
in the description of the aggregation process and 
understanding the significance of the runaway growth. However it 
is also likely that new experimental 
work to describe cloud droplet aggregation~\cite{Siebert:15}, 
and new observational discoveries on exoplanets \cite{Win+15}
will 
provide the missing insights which are required to complete the picture. 

\section*{Disclosure statement}

The authors are not aware of any affiliations, memberships,
funding, or financial
holdings that might be perceived as affecting the objectivity of this review.

\section*{Acknowledgements}

The work presented here was largely inspired from our collaborations with
G. Falkovich (AP) and B. Mehlig (MW), to whom we are particularly indebted.
We have also
benefited from discussions with J. Bec, E. Bodenschatz, G. Bewley, M. Bourgoin,
C. Clement, L. Collins, C. Connaughton, K. Gustavsson, E. Leveque,  
S. Malinowski, E. Meneguz, A. Naso, J. F. Pinton, M. Reeks, R. Shaw, 
M. Vo\ss kuhle and H. Xu.
Financial support from the french ANR (contract TEC2) is gratefully 
acknowledged. MW benefitted from visiting the Kavli Institute for Theoretical 
Physics, Santa Barbara, where this research was supported in part by the 
National Science Foundation under Grant No. NSF PHY11-25915.

\section{Appendix A: One-dimensional model for clustering and caustics}
\label{sec: 6}
 
In view of the role played by clustering effects and caustic
formation, it is desirable to see both of these effects in operation
in  a simplified, analytically tractable model for particle motion 
in a random flow. This appendix describes an exactly solvable
one-dimensional model \cite{Wil+03} which 
explains why the particle distribution
is a fractal \cite{Wil+10}, and why equation (\ref{eq: 4.3.2}) is 
a good model for the rate 
of formation of caustics. 
While the model is only exactly solvable in one dimension, the qualitative 
predictions are easily extended to higher dimensions.

\subsection{A one-dimensional model}
\label{sec: 6.1}

An incompressible flow in one dimension has no spatial variation, and
cannot generate either clustering or caustics. We will consider a compressible
one-dimensional model:
\begin{equation}
\label{eq: 6.1}
\dot x = v ~; ~~
\dot v = \gamma [u(x,t)-v]
\end{equation}
where  $\gamma $ is the inverse of the particle characteristic time, and 
$u(x,t)$ is  a random gaussian velocity field, delta-correlated in time, with  smooth spatial correlations:
\begin{equation}
\label{eq: 6.2}
\langle u(x,t)\rangle = 0 ~; ~~
~\nonumber \\
~%
\langle u(x,t)u(x',t')\rangle = C(x-x')\delta (t-t')
\end{equation}
\subsection{Preferential concentration and caustics: a general argument}
\label{sec: 6.2}

To understand both clustering and caustics, we write the equation of 
motion for small separations in positions, $\delta x$, and velocity $\delta v$ 
obtained by linearizing~Eq.(\ref{eq: 6.1}):
\begin{equation}
\label{eq: 6.3}
\delta \dot x = \delta v~; ~~
\delta \dot v = -\gamma \delta v+f(t)\delta x
\end{equation}
where $f(t) = \gamma \frac{\partial u}{\partial x}(x(t),t)$. 
Now consider the change of variable from $(\delta x,\delta v)$ to
$(Y,Z)$, defined by 
$Y={\rm ln}~\delta x$ and $ Z=\frac{\delta v}{\delta x}$.
In terms of these variables, the linearised equations of motion (\ref{eq: 6.3}) become:
\begin{eqnarray}
\label{eq: 6.6}
\dot Y & = & Z \label{eq: 6.6} \\
\dot Z & = & -\gamma Z+Z^2+f(t) \label{eq: 6.7}
\end{eqnarray}
The variable $Z$ obeys a stochastic differential equation, 
independent of $Y$, with 
$f$ acting as a forcing term. The 
variable $Y$ 
is then determined by a very simple evolution equation, which can 
interpreted as a generalised random walk. The evolution of $Y$ is then 
characterised by a drift velocity and a diffusion coefficient.  
The equation describing the evolution of $Y$ is invariant  under the
transformation $T_{\delta}: Y \rightarrow Y + \delta $, so 
the probability density of $Y$ must also be an eigenfunction of 
$T_{\delta }$.
This implies that  a steady-state probability density for $Y$
must be of the form $P(Y) = A \exp ( \alpha Y)$, which in terms of 
the original variable $\delta x$, leads to:
\begin{equation}
\label{eq: 6.10}
P(\delta x)=\delta x^{\alpha-1}
\end{equation}
The linearized equation of motion 
(\ref{eq: 6.6},\ref{eq: 6.7}) 
is only valid when $\delta x$ 
is very small, corresponding to sufficiently negative values of
$Y$. The solution (\ref{eq: 6.10}) corresponds to a normalisable distribution 
only if $\alpha>0$.

Equation~(\ref{eq: 6.10}) predicts that the expected number of particles 
in a ball of radius $\epsilon$ surrounding
a test particle is $\langle {\cal N}(\epsilon)\rangle \sim \epsilon^{D_2}$, 
where $D_2$ is the correlation dimension \cite{Wil+10}. The corresponding probability 
density is $P(\delta x)\sim \delta x^{D_2-1}$, so using (\ref{eq: 6.10})
leads to the conclusion that $D_2=\alpha$. This simple
argument, which is equally valid in higher dimensions, 
establishes why particles cluster onto a fractal set. 
 
Crucial to the statistical properties of $\delta x$ is the Lyapunov
exponent, $\lambda$, defined by:
\begin{equation}
\label{eq: 6.12}
\lambda=\lim_{t\to \infty}\frac{1}{t}{\rm ln}\left(\frac{\delta x(t)}{\delta x(0)}\right)
\end{equation}
where $\delta x(t)$ is the infinitesimal separation of two trajectories.
It follows that 
\begin{equation}
\label{eq: 6.13}
\lambda=\langle Z\rangle
\end{equation}
The variable $Z$ also has a clear interpretation in relation to the 
existence of caustics. Caustic singularities correspond to points where 
$\delta x=0$, 
so that at caustics $|Z|\to +\infty$. The 
nonlinear term in (\ref{eq: 6.7}) for $Z$ leads to a
finite time singularity, with $Z \to + \infty$. 
The trajectory returns with $Z=-\infty$ after the singularity.

In summary,  the equation of 
evolution  (\ref{eq: 6.7}) for $Z(t)$ provides important information,
concerning both preferential
concentration (effectively the exponent $\alpha$ in (\ref{eq: 6.10})),
as well as caustics formation, which
is the rate of formation of singularities of $Z$.

\subsection{Solution via a Fokker-Planck equation}
\label{sec: 6.3}

The fractal clustering and rate of caustic formation can both
be analysed for the one-dimensional model (\ref{eq: 6.1}), with the 
velocity field which is white-noise in time, (\ref{eq: 6.2}). 
Here, we concentrate upon the 
calculation of the rate of caustic formation, in order to show that 
this has a singular behaviour, analogous to (\ref{eq: 4.3.2}), in the limit
where inertial effects are negligible.
The Lyapunov exponent, as well as the correlation dimension can be accurately 
determined for this model~\cite{Wil+03}.

In our model, the stochastic term $f$ in (\ref{eq: 6.7}) is also
white-noise, with zero mean, and a variance $\langle f(t) f(t') \rangle
= 2 D \delta (t - t')$, the diffusion coefficient $D$ being simply deduced from the correlation tensor of $u$:
\begin{equation}
\label{eq: 6.15}
D=-\frac{1}{2}\frac{\partial^2 C}{\partial x^2}(0)
\ .
\end{equation}
The probability density for $Z$ is determined by a Fokker-Planck equation \cite{Wil+03}
\begin{equation}
\label{eq: 6.16}
\frac{\partial P}{\partial t}=
\frac{\partial {\cal J}}{\partial Z} ~ \mbox{     with     } ~ 
{\cal J}=( \gamma Z+Z^2 )P+ D \frac{\partial P}{\partial Z} 
\end{equation}
This is in the form of a continuity equation. Steady-state solution
in one dimension correspond to a
uniform flux ${\cal J} = J$, which determines the rate
of escape of $Z(t)$ to infinity, hence to the formation of caustics.
The search of a solution is facilitated by introducing a dimensionless variable $z$, 
a dimensionless parameter $\varepsilon$, and a potential 
$\phi(x,\varepsilon)$ 
\begin{equation}
\label{eq: 6.17a}
Z=\sqrt{\frac{D}{\gamma}z}
\ ,\ \ \ 
\varepsilon =\sqrt{\frac{D}{\gamma^3}}
\ ,\ \ \ 
\phi(z,\varepsilon)=\frac{z^2}{2} + \varepsilon \frac{z^3}{3}
\end{equation}
In terms of the dimensionless variables,
the distribution of the scaled variable $z$ is \cite{Wil+03}
\begin{equation}
\label{eq: 6.18}
P(z)= \frac{J}{\varepsilon\gamma^2}  \exp [ -\phi(z,\varepsilon)] 
\int_{-\infty}^z \exp[\phi(z',\varepsilon)]\ {\rm d}z' 
\end{equation}
Imposing the normalization condition for the probability density leads to 
an expression 
for the rate of caustic formation, $J$. In the limit as 
$\varepsilon \to 0$, the 
integral
over $x'$ is approximately $\sqrt{2\pi}\exp(1/6 \varepsilon^2)$ when $z$ is in the 
interval $[-1/\varepsilon,1/2\varepsilon]$. Integrating over $Z$ to normalise the distribution
yields the approximation
\begin{equation}
\label{eq: 6.19}
J= \frac{\gamma}{2\pi}\exp\left[-\frac{1}{6\varepsilon^2}\right]
\end{equation}
which is valid in the limit as $\varepsilon\to 0$ \cite{Gus+13}. This calculation supports the 
hypothesis that the rate of caustic formation has a non-analytic behaviour as 
the Stokes number approaches zero.



\end{document}